\documentstyle[aps,prc,preprint]{revtex}
\draft
\tighten
\title
{PARTIAL LEVEL DENSITIES FOR NUCLEAR DATA CALCULATIONS}
\author{M. Avrigeanu\setcounter{footnote}{3}\footnote{
	e-mail: mavrig@roifa.ifa.ro, fax: 401-4231701} and V. Avrigeanu}
\address{Institute for Physics and Nuclear Engineering "Horia 
         Hulubei",\\ P.O. Box MG-6, 76900 Bucharest, Romania}
\begin{document}
\maketitle
\vspace*{1.0in}

\noindent
{\bf Abstract}\\

The main formalisms of partial level densities (PLD) used in 
preequilibrium nuclear reaction models, based on the equidistant 
spacing model (ESM), are considered. A collection of FORTRAN77 
functions for PLD calculation by using 14 formalisms for the related 
partial-state densities is provided and 28 sample cases (73 versions) 
are described. The results are given in graphic form too. Composite 
(recommended) formulas, which include the optional use of various 
corrections, i.e. the advanced pairing and shell correction in 
addition to the Pauli effect, and average energy-dependent 
single-particle level densities for the excited particles and holes, 
are also given. The formalism comprises the density of particle-hole 
bound states, and the effects of an exact correction for the 
Pauli-exclusion principle are considered.\\

\newpage

{\bf PROGRAM SUMMARY}\\

\noindent
{\it Title of program:} PLD

\noindent
{\it Catalogue identifier:}

\noindent
{\it Program obtainable from:} CPC Program Library, Queen's University
of Belfast, N. Ireland

\noindent
{\it Licensing provisions:} none

\noindent
{\it Computer for which the program is designed and others on which it
is operable:} PCs (486 and Pentium)

\noindent
{\it Operating systems under which the program has been tested:} DOS

\noindent
{\it Programming language used:} FORTRAN 77 (MS-FORTRAN v.5.0)

\noindent
{\it Memory required to execute with typical data:} 491 Kbytes

\noindent
{\it No. of bytes in distributed program, including test data, etc.:} 
		1,256,659

\noindent
{\it Distribution format:} ASCII


\noindent
{\it Nature of physical problem:} \\
This Fortran code is a collection of subroutines for calculation of 
the partial nuclear level densities (PLD) mainly used in 
preequilibrium nuclear reaction models, by using 14 formalisms for the 
related partial state densities (PSD).\\

\noindent
{\it Method of solution:}\\
The main approaches to the calculation of the partial state density, 
based on the equidistant spacing model (ESM), are used. Composite 
(recommended) formulas including optionally various corrections, i.e. 
the advanced pairing and shell correction in addition to the Pauli 
effect, and average energy-dependent single-particle level (s.p.l.) 
densities for the excited particles and holes, are also involved. The 
density of the particle-hole bound states is moreover comprised, and 
the effects of an exact correction for the Pauli-exclusion principle  
are considered.\\

\noindent
{\it Restrictions on the complexity of the problem:}\\
Although the quantum-mechanical s.p.l. density and the {\it continuum 
effect} can also be reproduced by a corresponding Fermi-gas formula, 
to be used accordingly within the average energy-dependent PSD in 
multistep reaction models, this effect is not included. The 
calculation of PLD with linear momentum, of first interest for 
modelling preequilibrium-emission angular distributions, is not 
available either.\\

\noindent
{\it Typical running time:}\\
The execution time is strongly problem-dependent: it is roughly 
proportional to both the number of the excitation energies and the 
exciton configurations considered in the calculation, while consistent 
differences arise when various PSD formalisms are used. Twenty-eight 
sample cases with 73 versions which require from 0.1 to 1661 s on a PC 
Pentium/166MHz are provided.\\

\noindent
{\it Unusual features of the program:}\\
The PSD functions have been optimized for their independent use, in 
order to provide tools for PSD/PLD users (see [5]). The related 
drawback is the increase in execution time, while a proper use would 
involve the calculation of some coefficients only once in the main 
program. Second, the PLD.FOR has been organized so that various 
formulas and versions may be tried as well as the comparison between 
their predictions.\\

\noindent
{\it References}\\
$[$1$]$ M. Avrigeanu and V. Avrigeanu, J. Phys. G {\bf 20} (1994) 613.\\
$[$2$]$ M. Avrigeanu, A. Harangozo, and V. Avrigeanu, Rom. J. Phys. 
	{\bf 41} (1996) 77.\\
$[$3$]$ V. Avrigeanu, A. Florescu, A. Sandulescu, and W. Greiner,
	Phys. Rev. C {\bf 52} (1995) 1765.\\
$[$4$]$ M. Avrigeanu, A. Harangozo, V. Avrigeanu, and A.N. Antonov, 
	Phys. Rev. C {\bf 54} (1996) 2538; 
	\hspace*{0.2in} {\it ibid.}  {\bf 56} (1997) 1633.\\ 
$[$5$]$ M. Avrigeanu, I. \c Ste\c tcu, and V. Avrigeanu, 
	{\it Development of data file with partial level 
	\hspace*{0.2in} densities for nuclear data calculations. Final 
	report for IAEA Research Contract No. 8886, 
	\hspace*{0.2in} 15.12.1995-15.12.1997}, 
	http://tndln1.ifa.ro/\~{}vavrig/absf8886.html .\\

{\bf LONG WRITE-UP}\\

\noindent
{\bf 1. Introduction}\\

The particle-hole excitations caused by the nuclear reactions which 
proceed through a number of nucleon-nucleon interactions are described 
within either the semiclassical models or the quantum-statistical 
theories of the preequilibrium emission (PE) 
\cite{gadioli92,feshbach92} by means of the particle-hole 
state densities. Basic approaches to the partial state density (PSD) 
consist in combinatorial calculations performed in the space of 
realistic shell-model single-particle levels (s.p.l.) 
\cite{williams73}. Lenske {\it et al.} \cite{lenske94} have already 
used them in order to connect in a consistent way the 
quantum-statistical theories of the multistep-direct (MSD) and 
multistep-compound (MSC) processes \cite{feshbach80}. However, the 
strong dependence on the s.p.l. basic set is the main of several 
shortcomings inherent in the method (e.g.,
\cite{reffo94,blann89,nishioka88,sato91} and references therein).

The equidistant spacing model (ESM) state density \cite{ericson60} 
including the effect of the Pauli-exclusion principle 
\cite{williams71} is still widely used, as well as the 
phenomenological s.p.l. density value $g\sim A/14$ MeV$^{-1}$. Basic 
developments of the Williams formula \cite{williams71} are due to 
B\v et\' ak and Dobe\v{s} \cite{betak76} including the 
nuclear-potential finite depth correction, Stankiewicz {\it et al.} 
\cite{stankiewicz85} and Oblo\v zinsk\' y \cite{oblozinsky86} who 
added the bound-state condition, and Fu \cite{fu84} and Kalbach 
\cite{kalbach8789} who included an advanced pairing correction. 
Additional studies along this line have involved exact 
Pauli-correction calculation \cite{zhang88,baguer89,mao93}. 
Kalbach \cite{kalbach75,kalbach81,kalbach85} also discussed different 
energy dependences of the s.p.l. spacings and pointed out the 
necessity to study this subject closely related to PE surface effects, 
due to the interdependence of the corresponding assumptions. 
PSD including different energy-dependences 
of the excited-particle and hole state densities has recently been 
used \cite{avrigeanu94a} in the geometry-dependent hybrid (GDH) model 
\cite{blann7273,blann83}. A similar attempt 
\cite{avrigeanu94b,avrigeanu95} has focused on the MSD and MSC 
processes in the framework of the Feshbach-Kerman-Koonin (FKK) theory
\cite{feshbach80}. An independent semiclassical analysis 
\cite{avrigeanu95a,avrigeanu96} has additionally justified the surface 
localization of the most important first nucleon-nucleon interaction 
within PE processes, and provided average quantities useful for the
corresponding PSD calculation \cite{avrigeanu97}.

The various PLD formalisms based on the ESM Williams-type formula, 
which are still extensively used in nuclear-reaction calculations, 
determined the need for an appropriate subroutine collection. It is on
the request of a project concerning a reference input-parameter 
library for nuclear model calculations \cite{oblozinsky94} that the 
present work is based. Thus, the program PLD.FOR is a collection of 
algorithms developed until now and widely used for PSD/PLD 
calculations. The one- and two-fermion system versions of six 
different approaches and one composite (recommended) formula including 
various corrections are available as FORTRAN77 functions. 

    The main points of the PSD and PLD formalisms are presented in 
Section 2. At the same time, the sample cases (Table I) for the
program PLD.FOR are described. The structure of the program and the
input-data description is given in Section 3. Finally, a worked 
example is presented in Section 4.\\

\noindent
{\bf 2. Formalism}\\

\noindent
{\it 2.1. Partial state density in the uniform spacing model}\\

The state density of a system of $p$ excited particles above the Fermi 
level and $h$ holes below it, considered within the {\it uniform 
spacing model} (based on a constant spacing $d$=1/$g$ between the 
non-degenerate single-particle levels) at the last occupied level in 
the ground state of the nucleus, was obtained by Williams 
\cite{williams71} 

\begin{equation} \label{eq:1}
   \omega(p,h,E)={{g^n(E-A)^{n-1}}\over{p!h!(n-1)!}} \: ,
\end{equation}
by decreasing the excitation energy $E$ with the correction for the 
Pauli blocking

\begin{equation} \label{eq:2}
   A={{p(p+1)+h(h-1)}\over{4g}}-{{h}\over{2g}}
\end{equation}
with respect to Ericson early formula \cite{ericson60} for $n$=$p+h$ 
total number of excitons. The sum of the partial state densities for 
all allowed particle-hole numbers $p$=$h$ is consistent with the total 
nuclear state density formula obtained in the frame of the ESM of the 
one-component Fermi gas \cite{ericson60}

\begin{equation} \label{eq:3}
 \omega_1(E)={{\mbox{exp}[2({{\pi^2}\over{6}}gE)^{1/2}]}\over
              {\sqrt{48}\,E}} \: .
\end{equation}
The asymptotic equality 

\begin{equation} \label{eq:4}
 \omega_1(E) \simeq \sum_{p=h} \omega(p,h,E) \: ,
\end{equation}
is illustrated in Fig. 1(a) (the sample cases 1/1A) as proved by 
Williams for the generic value $g$=1 MeV$^{-1}$ of the s.p.l. density. 

Similarly, in the case of the two kinds of fermions considered, with 
the $g_{\pi}$ and $g_{\nu}$ being the single-proton and single-neutron 
state densities, respectively, the PSD for $p_{\pi}$ and $h_{\pi}$ 
proton particle and hole numbers, respectively, and $p_{\nu}$ and 
$h_{\nu}$ neutron particle and hole numbers  
($n=p_{\pi}+h_{\pi}+ p_{\nu}+h_{\nu}$) is

\begin{equation} \label{eq:5}
   \omega(p_{\pi},p_{\nu},h_{\pi},h_{\nu},E)=
    {{g_{\pi}^{p_{\pi}+h_{\pi}}g_{\nu}^{p_{\nu}+h_{\nu}}(E-B)^{n-1}}
    \over{p_{\pi}!p_{\nu}!h_{\pi}!h_{\nu}!(n-1)!}} \: ,
\end{equation}
where the Pauli effect correction has now the form

\begin{equation} \label{eq:6}
 B={{1}\over{4}} \left[
      {{p_{\pi}(p_{\pi}+1)+h_{\pi}(h_{\pi}-1)}\over{g_{\pi}}}
     +{{p_{\nu}(p_{\nu}+1)+h_{\nu}(h_{\nu}-1)}\over{g_{\nu}}}\right]
     -{{1}\over{2}}\left({{h_{\pi}}\over{g_{\pi}}}
     +{{h_{\nu}}\over{g_{\nu}}}\right) \: .
\end{equation}

  The corresponding ESM total nuclear state density for a two-fermion 
system with an average total single-particle state density 
$g$=$g_{\pi}$+$g_{\nu}$ and related level density parameter 
$a$=$(\pi^2/6)g$ \cite{ericson60}

\begin{equation} \label{eq:7}
 \omega_2(E)={{\sqrt{\pi}}\over{12}}
             {{\mbox{exp}[2(aE)^{1/2}]}\over{a^{1/4}\,E^{5/4}}}
\end{equation}
is also consistent with the sum of the partial state densities 
(\ref{eq:5}) for all allowed pairs of particle-hole numbers 
$p_{\pi}$=$h_{\pi}$ and $p_{\nu}$=$h_{\nu}$, i.e.

\begin{equation} \label{eq:8}
 \omega_2(E) \simeq \sum_{p_{\pi}=h_{\pi},p_{\nu}=h_{\nu}}
   \omega(p_{\pi},p_{\nu},h_{\pi},h_{\nu},E) \: ,
\end{equation}
shown in Fig. 1(b) (Cases 2/2A) for the similar generic values 
$g_{\pi}$=$g_{\nu}$=$g$/2=1 MeV$^{-1}$. It results that 
Eq. (\ref{eq:8}) is true within 4\% for $E >$ 3 MeV. A comparison of 
the PSD given by the Williams one- and two-fermion formulas for the 
real case of the nucleus $^{93}$Nb is carried out in Fig. 1(c) (Cases
3/3A) by using the phenomenological value $g$=$A/13$ MeV$^{-1}$ and 
the derived quantities

\begin{mathletters}
\begin{eqnarray} \label{eq:9}
 g_{\pi} & = & {{Z}\over{A}}g \\
 g_{\nu} & = & {{A-Z}\over{A}}g \: .
\end{eqnarray}
\end{mathletters}
The renormalization of the PSDs given by one-component Williams-type 
formula can be done by using the ratio between the total state 
densities given by the two- and respectively one-component Fermi gas 
formulas of the general form \cite{gruppelaar83,akkermans85} 

\begin{equation} \label{eq:10}
 \omega_1(E)={{\mbox{exp}[2(aU)^{1/2}]}\over{\sqrt{48}\,U}}
\end{equation}

\begin{equation} \label{eq:11}
 \omega_2(E)={{\sqrt{\pi}}\over{12}}
        {{\mbox{exp}[2(aU)^{1/2}]}\over{a^{1/4}\,(U+t)^{5/4}}} \: ,
\end{equation}
where the nuclear temperature $t$ corresponding to the effective 
excitation energy $U$ (see below) is defined by \cite{dilg73}

\begin{equation} \label{eq:12}
 U=at^2 - t \: .
\end{equation}
Thus, the renormalized one-fermion PSD has been defined by

\begin{equation} \label{eq:13}
\omega(n,E)=f(U) \, \omega_1(p,h,E) \: ,
\end{equation}
where the renormalization factor $f(U)$ (two-fermion system correction
-- TFC)

\begin{equation} \label{eq:14}
 f(U)=\left({{\pi}\over{3}}\right)^{1/2}{{U}\over{a^{1/4}(U+t)^{5/4}}}
\end{equation}
has only a weak energy dependence approximately equal to $U^{-1/4}$. 

     Since each of the closed formulas (\ref{eq:10}) and
(\ref{eq:11}) are asymptotically equal to the sum of the one- and
respectively two-component PSDs over all allowed particle-hole numbers 
$p$=$h$, the corresponding sum of the renormalized PSD (\ref{eq:13}) 
is consistent with the closed formula for a two-fermion system, as 
shown in Fig. 1(d) (Cases 3/3A/3B).\\ 

\noindent
{\it 2.2. Bound-state and finite depth corrections}\\

  The limitation on the hole maximum energy due to the finite depth 
of the nuclear potential \cite{betak76} as well as on the maximum 
particle excitation by the nucleon binding energy $B$ in the case of 
the bound states, yielded the approximate one-fermion ESM formula 
\cite{oblozinsky86}

\begin{equation} \label{eq:15}
 \omega(p,h,E)={{g^n}\over{p!h!(n-1)!}}\sum_{i=0}^p
 \sum_{j=0}^h\,(-1)^{i+j}\,C_p^i\,C_h^j(E-A_{ph}-iB-jF)^{n-1}
 \Theta(E-\alpha_{ph}-iB-jF) \: ,
\end{equation}
where $F$ is the Fermi energy which is now considered to be halfway 
between the last filled and the first free s.p.l. in the nucleus 
ground state \cite{oblozinsky86,kalbach75} in order to have a PSD form 
symmetrical in $p$ and $h$. Under these circumstances, the Pauli 
correction term in Eq. (\ref{eq:15}) is

\begin{equation} \label{eq:16}
   A_{ph}={{p(p-1)+h(h-1)}\over{4g}} \: ,
\end{equation}
while

\begin{equation} \label{eq:17}
  \alpha_{ph} = {{p^2+h^2}\over{2g}}
\end{equation}
is the minimum energy of the $(p,h)$ state due to the Pauli blocking.
$\Theta$ in Eq. (\ref{eq:15}) is the unit step function, i.e. 1 for a
positive argument and 0 otherwise. The effect of the nuclear-potential 
finite depth on the PSD as firstly pointed out by B\v et\' ak and 
Dobe\v{s} \cite{betak76}, and the additional one due to the condition 
of bound states, included by Oblo\v zinsk\' y \cite{oblozinsky86}, are
shown in Fig. 2 (Case 4) and Fig. 3(a) (Cases 5/5A/5B). The changes 
obtained by releasing consecutively the finite-depth potential and 
bound state conditions, by means of large $F$- and $B$-values, are 
shown in the later case for the basic exciton configuration $1p1h$. 
The consequences of these conditions on the total nuclear state 
density, as well as the corresponding results obtained by using the
asymptotic formula (\ref{eq:3}), are illustrated in Fig. 3(b) (Cases
6/6A/6B).

   The same method \cite{oblozinsky86} applied within the two-fermion 
ESM led to the density of partial two-fermion bound states of the form

\[
 \omega(p_{\pi},h_{\pi},p_{\nu},h_{\nu},E)=
    {{g_{\pi}^{p_{\pi}+h_{\pi}}g_{\nu}^{p_{\nu}+h_{\nu}}}\over
     {p_{\pi}!h_{\pi}!p_{\nu}!h_{\nu}!(n-1)!}}
     \sum_{i_{\pi}=0}^{p_{\pi}}
     \sum_{i_{\nu}=0}^{p_{\nu}}
     \sum_{j_{\pi}=0}^{h_{\pi}}
     \sum_{j_{\nu}=0}^{h_{\nu}}
      (-1)^{i_{\pi}+i_{\nu}+j_{\pi}+j_{\nu}}\,
      C_{p_{\pi}}^{i_{\pi}}\,
      C_{p_{\nu}}^{i_{\nu}}\,
      C_{h_{\pi}}^{j_{\pi}}\,
      C_{h_{\nu}}^{j_{\nu}}\, 
\]
\[
 \times(E-A_{p_{\pi}h_{\pi}p_{\nu}h_{\nu}}
 -i_{\pi}B_{\pi}-i_{\nu}B_{\nu}-j_{\pi}F_{\pi}-j_{\nu}F_{\nu})^{n-1}
\]
\begin{equation} \label{eq:18}
 \times \Theta(E-\alpha_{p_{\pi}h_{\pi}p_{\nu}h_{\nu}}-i_{\pi}B_{\pi}-
         i_{\nu}B_{\nu}-j_{\pi}F_{\pi}-j_{\nu}F_{\nu}) \: ,
\end{equation}
where

\begin{equation} \label{eq:19}
   A_{p_{\pi},h_{\pi},p_{\nu},h_{\nu}}={{1}\over{4}} \left[
      {{p_{\pi}(p_{\pi}-1)+h_{\pi}(h_{\pi}-1)}\over{g_{\pi}}}
     +{{p_{\nu}(p_{\nu}-1)+h_{\nu}(h_{\nu}-1)}\over{g_{\nu}}}\right]
\end{equation}
and

\begin{equation} \label{eq:20}
   \alpha_{p_{\pi},h_{\pi},p_{\nu},h_{\nu}}=
     {{p_{\pi}^2+h_{\pi}^2}\over{2g_{\pi}}} +
     {{p_{\nu}^2+h_{\nu}^2}\over{2g_{\nu}}}
\end{equation}
are symmetrical in the respective particle and hole numbers.
The total nuclear state density, i.e. the sum of all PSD given by 
Eq.(\ref{eq:5}) for allowed pairs of proton and neutron particle-hole 
numbers, calculated with/without the potential finite-depth and 
bound-state conditions are shown in Fig. 3(c) (Cases 7/7A/7B).\\

\noindent
{\it 2.3. The advanced pairing correction}\\

\noindent
{\it 2.3.1. One-component Fermi gas formula}\\

(a). In order to take into account the pairing interaction between 
nucleons a correction was included \cite{fu84} into the formula 
derived by Williams in the frame of the free Fermi-gas model (FGM), 
function of the particle and hole numbers as well as excitation energy 
of the configuration. The Pauli correction was also modified to be 
consistent with the pairing  correction, so that the PSD formula 
became

\begin{equation} \label{eq:21}
   \omega(p,h,E,P+B)={{g^n(E-P-B)^{n-1}}\over{p!h!(n-1)!}} \: ,
\end{equation}
where $B$ is the modified Pauli correction following the Williams term 
$A$:

\begin{equation} \label{eq:22}
B=A[1+(2g\Delta/n)^2]^{1/2} \: ,
\end{equation}
and the pairing correction term

\begin{equation} \label{eq:23}
    P = {{1}\over{4}}g(\Delta_0^2-\Delta^2)
\end{equation}
is determined by the ground- and excited-state gaps $\Delta_0$ and 
$\Delta(p,h,E)$. The former is related to the condensation energy 
$C=g\Delta_0^2/4$. On the other hand, the constant pairing correction 
$U_p$ for the total state density, based on the odd-even mass 
differences (e.g., \cite{dilg73}), may be rather well related to the 
value $P(\hat n)$ \cite{fu84} where $\hat n$ is the most probable exciton 
number. Since $\Delta$=0 if $n\ge\hat n$, it results that $\Delta_0$ can 
be derived from the relation $U_p=g\Delta_0^2/4$. Then, Fu obtained 
the following parametrizations for $\Delta$ \cite{fu84,kalbach8789}:

\begin{eqnarray} \label{eq:24}
 {{\Delta}\over{\Delta_0}} & = & 0.996-1.76(n/n_c)^{1.60}(E/C)^{-0.68}
 \hspace*{.25in}\mbox{if} \: \: E \geq E_{phase} \nonumber \\
         & = & 0 \hspace*{2.48in}\mbox{if} \: \: E < E_{phase} \: ,
\end{eqnarray}
where $n_c=0.792g\Delta_0$ is the critical number of excitons and
$E_{phase}$ is the energy of the pairing phase transition given by

\begin{eqnarray} \label{eq:25}
 E_{phase} & = & C \, [0.716+2.44(n/n_c)^{2.17}]
    \hspace*{.25in} \mbox{if} \: \: n/n_c \geq 0.446  \nonumber \\
          & = & 0 \hspace*{1.95in} \mbox{if} \: \: n/n_c < 0.446 \: .
\end{eqnarray}
Actually, the lower limit in Eq. (\ref{eq:25}) was adopted 
\cite{kalbach8789} in order to take into account explicitly the lack 
of a phase transition for small $n$. The original comparison of the 
calculated PSD with and without pairing correction is shown in Fig. 
4(a) (Case 8, with the threshold-condition released). One should note 
that the respective equations provide PSD-values even below the 
minimum excitation energies (thresholds) characteristic of each 
configuration 

\begin{eqnarray} \label{eq:26}
 U_{th} & = & C \, [3.23(n/n_c)-1.57(n/n_c)^2]
      \hspace*{.25in} \mbox{if} \: \: n/n_c \leq 0.446  \nonumber \\
 &=&C\, [1+0.627(n/n_c)^2]\hspace*{.84in}\mbox{if} \: \: n/n_c > 0.446 
 \: .
\end{eqnarray}
The correct densities, i.e. above the respective thresholds, are shown
in Fig. 4(b) (Case 8).  

  The total state density obtained as the sum of all PSD for allowed 
pairs of particle-hole numbers $p=h$ are compared with the closed
formula of the one-component Fermi gas used at an effective excitation 
energy decreased by the constant pairing correction $U_p$ \cite{fu84}:

\begin{equation} \label{eq:27}
\omega_1(E,U_p)={{\mbox{exp}\{2[a(E-U_p)^{1/2}]\}}\over
                 {\sqrt{48}\,(E-U_p)}} \: .
\end{equation}
A distinct underestimation by this formula of the sum of PSD for 
$E<$8-10 MeV also results from the comparison shown in Fig. 4(b) (Case 
8) for the constant pairing correction $U_p$=3.5 MeV. The analysis 
performed with respect to the effect of various $U_p$ values on this 
agreement \cite{fu84} is shown in Fig. 4(c) (Case 9) for the generic 
values $g$=4 MeV$^{-1}$ and 0, 2, and 4 MeV for $U_p$.

The relationship between the pairing corrections for the PSD and the 
ESM total state density was extended to considering the nuclear-shell 
effects by using an additional back-shift $S$ of the effective 
excitation energy \cite{fu84}

\begin{equation} \label{eq:28}
   \omega(p,h,E,P+B+S)={{g^n(E-P-B-S)^{n-1}}\over{p!h!(n-1)!}} \: .
\end{equation}
By analogy with the BSFG model \cite{dilg73} for the two-fermion 
system total state density

\begin{equation} \label{eq:29}
 \omega_2(E)={{\sqrt{\pi}}\over{12}}
             {{\mbox{exp}\{2[a(E-\Delta)]^{1/2}\}}\over
              {a^{1/4}\,(E-\Delta+t)^{5/4}}} \: ,
\end{equation}
the back-shift parameter is connected to the BSFG virtual 
ground-state shift parameter $\Delta$ through the relation

\begin{equation} \label{eq:30}
  \Delta=U_p \, + \, S 
\end{equation}
Thus, the predictions of the closed formula of the one-component Fermi 
gas (\ref{eq:27}) for the effective excitation energy $U=E-\Delta$, 
and the total state density standing for the PSDs given by Eq. 
(\ref{eq:28}), should be asymptotically equal as shown by the upper 
curves in Fig. 4(d) (Cases 10/10A). The lack of consistency between 
the sum of the one-component PSDs and the two-fermion BSFG state 
density formula has been illustrated by means of the various 
predictions of one- and two-component closed formulas with the same 
parameters for the excited odd-$A$ nucleus $^{41}$Ca \cite{fu84}, also 
shown in Fig. 4(d). 

(b). An improved implementation of the advanced pairing correction 
within the Williams formula

\begin{equation} \label{eq:31}
   \omega(p,h,E,A_K)={{g^n[E-A_K(p,h)]^{n-1}}\over{p!h!(n-1)!}}
\end{equation}
adopted a Pauli correction function symmetric in particles and holes,
which also included the effects of a passive hole \cite{kalbach8789}

\begin{equation} \label{eq:32}
   A_K(p,h)= E_{thresh}(p,h)-{{p(p+1)+h(h+1)}\over{4g}} \: ,
\end{equation}
where $E_{thresh}(p,h)=p_m^2/g$ and $p_m=maximum(p,h)$. The inclusion 
of the pairing interaction led to the modified form of the threshold 
energy for a given exciton configuration

\begin{equation} \label{eq:33}
E_{thresh}(p,h)= {{g(\Delta_0^2-\Delta^2)}\over{4}}+
     p_m\left[\left({{p_m}\over{g}}\right)^2+\Delta^2\right]^{1/2}
\end{equation}
A third term was added to the modified Pauli-and-pairing correction 

\begin{equation} \label{eq:34}
A_K(p,h)= E_{thresh}(p,h)-{{p(p+1)+h(h+1)}\over{4g}}+
           {{(p-1)^2+(h-1)^2}\over{gF(p,h)}} \: ,
\end{equation}
where

\begin{equation} \label{eq:35}
F(p,h)  = 12+4g[E-E_{thresh}(p,h)]/p_m \: ,
\end{equation}
in order to obtain the PSD values of $g$ and $2g$ for 
$E=E_{thresh}(p,h)$ and $E=E_{thresh}(p,h)+1/g$, respectively. The 
comparison of the values given by the formulas of Fu \cite{fu84} and 
Kalbach \cite{kalbach8789} above $E_{thresh}(p,h)$ for each exciton 
configuration, respectively, is reproduced in Fig. 5(a) (Case 12). 

The pairing effects were also included in the total state-density 
formula by using an effective excitation energy decreased by the 
effective pairing shift \cite{kalbach8789}

\begin{equation} \label{eq:36}
 P_{eff}(E,C)=maximum(E_2,C/\{1+\mbox{exp}[4(0.625-E/C)]\}) \: ,
\end{equation}
where the quantity $E_2$ is written as 

\begin{eqnarray} \label{eq:37}
 E_2 &=& C \, [1+2.508/(n_c)^2] 
       \hspace*{1.2in} \mbox{if} \: \: n_c \leq 4.48   \nonumber \\
     &=& C \, [6.46/n_c-6.28/(n_c)^2] \hspace*{.84in} 
                                   \mbox{if} \: \: n_c \geq 4.48 \: . 
\end{eqnarray}
The sum of the PSD provided by Eq. (\ref{eq:36}) is now consistent 
with the Fermi-gas formula (\ref{eq:37}) provided that the constant 
$U_p$ is replaced by the effective-energy shift $P_{eff}(E,C)$, as
shown in Fig. 5(b) (Case 13).

(c). The PSD formula also within the ESM and based on an exact 
calculation of the Pauli correction term \cite{baguer89}, extended to 
the case of the finite well depth and bound states, and including the 
Kalbach \cite{kalbach8789} pairing correction, is \cite{mao93}

\[
 \omega(p,h,E)={{g^n}\over{p!h!}}\sum_{i=0}^p\sum_{j=0}^h\,(-1)^{i+j}
               \,C_p^i\,C_h^j \sum_{\lambda=0}^{n-1}
               (E-E_{thresh}-iB-jF)^{n-1-\lambda}
\]
\begin{equation} \label{eq:38}
               \times\Theta(E-E_{thresh}-iB-jF)B(p,h,\lambda)
               {{1}\over{(n-1-\lambda)!}} \: ,
\end{equation}
where the coefficients $B(p,h,\lambda)$ have the expression 

\begin{equation} \label{eq:39}
 B(p,h,\lambda)=\sum_{\lambda_1}^{\lambda}
                 C(p,\lambda_1)C(h,\lambda-\lambda_1) \: , 
\end{equation}
and the coefficients $C(m,\lambda)$ are determined by the recursive
relation 

\begin{equation} \label{eq:40}
 C(m,\lambda)=\sum_{i=0}^{\lambda}{{1}\over{i!}}
              b_i(-m/g)^iC(m-1,\lambda-i) \: ,
\end{equation}
with

\begin{eqnarray} \label{eq:41}
C(0,\lambda)&=&1\hspace*{.25in} \mbox{for} \: \: \lambda=0 \nonumber\\
            &=&0\hspace*{.25in} \mbox{for} \: \: \lambda \neq 0 \: ,
\end{eqnarray}
while the Bernouli numbers $b_i$ can be found in mathematical tables. 
Eq. (\ref{eq:38}) becomes the PSD expression of Baguer \cite{baguer89}
in the limiting case of large $B$ and $F$. The exact coefficient 
$B(p,h,\lambda)$ corresponds to the factor

\begin{equation} \label{eq:42}
 {{1}\over{\lambda!}}
     \left({{p^2+p}\over{4g}}+{{h^2+h}\over{4g}}\right)^\lambda \: ,
\end{equation}
which follows from the approximate formula of Oblo\v{z}insk\'{y}
\cite{oblozinsky86} if the energy-term power is expanded by means of
the binomial theorem. The comparison with the PSD values obtained by 
means of Kalbach formula and parameters $g$=14 MeV$^{-1}$ and 
$\Delta_0$=1 MeV, shown in Fig. 5(c) (Case 17A), proves the 
suitability of the Pauli-correction approximation. At the same time, 
the total state density given by the sum of the corresponding PSDs 
calculated by using the parameter global values $g$=8 MeV$^{-1}$ and 
$\Delta_0$=1 MeV \cite{oblozinsky86,mao93} is in good agreement with 
the one-fermion closed formula (\ref{eq:42}) by using the 
effective-energy shift $P_{eff}(E,C)$ [Fig. 5(d) and Case 17].

The comparison of Eq. (\ref{eq:38}) with the Oblo\v{z}insk\'{y} 
formula can be carried out by neglecting the pairing effect, i.e. 
through substitution of the threshold energy $E_{thresh}$ by the Pauli 
energy $\alpha_{ph}$. While Oblo\v{z}insk\'{y} results for small 
exciton numbers are well represented (Case 15), a relative deviation 
of about 10-40\% exists between the two sets of calculated PSDs when 
larger $2p3h$ and $3p2h$ exciton numbers are involved [Fig. 6(a) and 
Case 15A]. On the other hand, the inclusion of the pairing effect 
\cite{mao93} decreases the partial bound-state densities for small $E$ 
and enhances them for large $E$ [Fig. 6(b) and Case 16].\\

\noindent
{\it 2.3.2. Two-component Fermi gas formula}\\

(a). The PSD formula derived by Williams in the frame of the 
two-component free Fermi-gas model, became by inclusion of the 
pairing correction $P_2$ \cite{fu91}

\begin{equation} \label{eq:43}
   \omega(p_{\pi},h_{\pi},p_{\nu},h_{\nu},E,P_2+B_2+S)=
   \left({ {g} \over {2} }\right)^n{ {(E-P_2-B_2-S)^{n-1}}
   \over {p_{\pi}!h_{\pi}!p_{\nu}!h_{\nu}!(n-1)!} } \: ,
\end{equation}
where $B_2$ is the simple extension of the one-fermion system 
correction factor for the Pauli-exclusion principle modified by the 
pairing effects. The pairing correction term has been adopted under 
the assumption of no pairing interaction between the protons and 
neutrons so that

\begin{equation} \label{eq:44}
P_2(E,n_{\pi},n_{\nu}) = P_1(E_{\pi},n_{\pi}) + P_1(E_{\nu},n_{\nu}) 
	\: ,
\end{equation}
where

\begin{mathletters}
\begin{eqnarray} \label{eq:45}
  P_1(E_{\pi},n_{\pi}) & = & {{1}\over{4}}g_{\pi}[\Delta_{0\pi}^2-
                             \Delta_{\pi}^2(E_{\pi},n_{\pi})] \: ,\\
  P_1(E_{\nu},n_{\nu}) & = & {{1}\over{4}}g_{\nu}[\Delta_{0\nu}^2-
                             \Delta_{\nu}^2(E_{\nu},n_{\nu})] \: ,
\end{eqnarray}
\end{mathletters}
and $E=E_{\pi}+E_{\nu}$, $n_{\pi}=p_{\pi}+h_{\pi}$, and 
$n_{\nu}=p_{\nu}+h_{\nu}$. Based on the pairing theory for two kinds 
of fermions, an approximate solution was adopted for the gaps 
$\Delta_{\pi}$ and $\Delta_{\nu}$. Actually, by using the mean 
gap-approximation, i.e. $g_{\pi}$=$g_{\nu}$ and 
$\Delta_{0\pi}$=$\Delta_{0\nu}$, it was shown that the results 
obtained for the one-fermion system can be used for each of the two 
systems, while the pairing theory yields that the proton system and 
the neutron system are excited isothermally \cite{fu91}. Next, the 
following simple procedure was chosen to define $E_{\pi}$ and 
$E_{\nu}$ to approximately 10\% of the exact values except for 
energies near the threshold

\begin{mathletters} \label{eq:46}
\begin{eqnarray} 
  E_{\pi} & = & En_{\pi}/n \: ,\\
  E_{\nu} & = & En_{\nu}/n \: ,
\end{eqnarray}
\end{mathletters}
with the overall error in $P_2$ estimated to be about 2\% except for 
energies near the threshold. Moreover, similar to the one-fermion 
case, the pairing correction $U_p$ for the total state density is
related to $P_2(E,\hat n_{\pi},\hat n_{\nu})$ \cite{fu91}

\begin{equation} \label{eq:47}
 U_p = P_2(E,\hat n_{\pi},\hat n_{\nu}) \: ,
\end{equation}
so that 

\begin{equation} \label{eq:48}
 \Delta_{0\pi}^2=\Delta_{0\nu}^2=4U_p/g \: ,
\end{equation}
i.e. the ground-state pairing gaps in the proton system and the 
neutron system are both equal to that of the one-fermion system if 
the same values of $g$ and $U_p$ are used. 

(b). The improved implementation of the pairing correction given by 
Kalbach \cite{kalbach8789} has the two-component version

\begin{equation} \label{eq:49}
   \omega(p_{\pi},h_{\pi},p_{\nu},h_{\nu},E,A_K)=
   {{g_{\pi}^{n_{\pi}} g_{\nu}^{n_{\nu}}
   [E-A_K(p_{\pi},h_{\pi},p_{\nu},h_{\nu})]^{n-1}}\over
   {p_{\pi}!h_{\pi}!p_{\nu}!h_{\nu}!(n-1)!}} \: ,
\end{equation}
where 

\begin{equation} \label{eq:50}
  A_K(p_{\pi},h_{\pi},p_{\nu},h_{\nu})=
  A_K(p_{\pi},h_{\pi}) + A_K(p_{\nu},h_{\nu}) \: ,
\end{equation}
and the respective one-fermion expressions are used except the 
function $F(p,h)$ which now has the form \cite{kalbach8789} 

\begin{equation} \label{eq:51}
F(p_i,h_i)  = 12n_i/n+4g_i[E_i-E_{thresh}(p_i,h_i)]/p_{im} \: ,
\end{equation}
where $i$ can be either $\pi$ or $\nu$. The average excitation
energies of the two kinds of nucleons have also been assumed to be 
proportional to the number of degrees of freedom of each type.

Additional comments should concern Kalbach \cite{kalbach8789} 
energy-dependent pairing corrections for the consistency between the 
total state-density closed formula of the two-component Fermi gas at
the effective excitation energy decreased by the effective-energy 
shift $P_{eff}(E,C)$, and the sum of all PSD provided by Eq. 
(\ref{eq:49}) for allowed pairs of particle-hole numbers 
$p_{\pi}$=$h_{\pi}$ and $p_{\nu}$=$h_{\nu}$, as shown in Fig. 7(a) 
(Case 13A). It was expected that the two-fermion PSDs sum should be 
well approximated by the Fermi gas formula when 
$P$=$C_{\pi}$+$C_{\nu}$, at excitation energies not too close to 
threshold, under the assumption of the same gap parameters 
$\Delta_{0\pi}$ and $\Delta_{0\nu}$. However, the use of the sum 
$P_{eff}(E,C_{\pi})$ and $P_{eff}(E,C_{\nu})$ as well as of the 
constant $P$=$U_p$ is shown in Fig. 7(b) (Case 13A) to be correct only 
for $E/C\geq$2. The PSDs sum is underestimated, i.e. the pairing 
correction is overestimated at the lowest energies where the exciton 
configurations of only one kind of fermions are significant 
[Fig. 7(a)]. This suggests the use at these energies of only one 
correction term out of the two $P_{eff}(E,C_{\pi})$ and 
$P_{eff}(E,C_{\nu})$. The gradual inclusion of the second one, by
means of the sum $P_{eff}(E,C_{\pi})$+$xP_{eff}(E,C_{\nu})$ with $x$
between 0 and 1 for $E/C$ varying from 1 to 2 as also shown in Fig. 
7(c), is correct just above the first threshold. Therefore, we found 
the following form was necessary to obtain the closed-formula 
predictions rather consistent with the PSDs sum as in Fig. 7(b) 

\begin{equation} \label{eq:52}
 P_{\pi+\nu,eff}(E,C)
= P_{eff}(E,C_i)\left({{E-E_2}\over{2C-E_2}}+1\right) \: ,
\end{equation}
where $i$ can be either $\pi$, or $\nu$ or an average of the two 
terms, while the quantity $E_2$ 

\begin{eqnarray} \label{eq:53}
 E_2 &=& C_i \, [1+2.508/(n_{ci})^2] 
      \hspace*{1.2in} \mbox{if} \: \: n_{ci} \leq 4.48 \nonumber \\
     &=& C_i \, [6.46/n_{ci}-6.28/(n_{ci})^2] \hspace*{.84in} 
                                \mbox{if} \: \: n_{ci} \geq 4.48 \: ,
\end{eqnarray}
is the ground-state threshold energy for the $n_i$=2 states, which 
Kalbach \cite{kalbach8789} also involved in the definition of 
$P_{eff}(E,C)$. The reduced consistency of the two curves just above 
$E$=$C$ could be shifted to the region above the threshold by 
replacing the energy $E_2$ with the zero value, i.e. by carrying out 
the transition described by Eq. (\ref{eq:52}) in the energy range 
between 0 and C. The corresponding pairing correction is denoted by 
$P'_{\pi+\nu,eff}$ in Fig. 7(b).

The comparison of the two-fermion PSDs sum with the sum of the 
one-fermion PSDs \cite{kalbach8789} multiplied by the two-fermion 
system correction \cite{fu86,akkermans85} for all allowed 
particle-hole numbers $p=h$, as well as of the corresponding 
closed-formula values, is shown in Fig. 7(d) (Cases 13A/13B). The 
agreement of the related quantities is quite good but only at 
$E/C\geq$2. The use of $P_{eff}(E,C)$ within the two-component 
Fermi-gas total state density, which is the case when the two-fermion 
system correction is used, leads once again to an underestimation 
around the condensation energy $C$. This could be the main limit of 
the TFC method, which reveals the need for a different pairing 
correction at energies $E/C\leq$2 within the complete two-fermion 
system approach.

(c). The PSD formula within the ESM and exact calculation of the Pauli 
correction term \cite{mao93}, extended to the case of the finite well 
depth and bound states and including the Kalbach \cite{kalbach8789} 
pairing correction, has the following form in the two-fermion system 
case 

\[
   \omega(p_{\pi},h_{\pi},p_{\nu},h_{\nu},E)=
 { {  g_{\pi}^{n_{\pi}}g_{\nu}^{n_{\nu}}     }\over
   {  p_{\pi}!h_{\pi}!p_{\nu}!h_{\nu}!(n-1)!  }  }
 \sum_{i_{\pi}=0}^{p_{\pi}}\, \sum_{j_{\pi}=0}^{h_{\pi}}\,
 \sum_{i_{\nu}=0}^{p_{\nu}}\, \sum_{j_{\nu}=0}^{h_{\nu}}\,
 (-1)^{i_{\pi}+j_{\pi}+i_{\nu}+j_{\nu}}\,
 C_{p_{\pi}}^{i_{\pi}}\, C_{h_{\pi}}^{j_{\pi}}\,
 C_{p_{\nu}}^{i_{\nu}}\, C_{h_{\nu}}^{j_{\nu}}\,
\]
\begin{equation} \label{eq:54}
 \sum_{\lambda=0}^{n-1}\,t^{n-1-\lambda}\theta(t)
 A(p_{\pi},h_{\pi},p_{\nu},h_{\nu},\lambda){{1}\over{(n-1-\lambda)!}} 
 \: ,
\end{equation}
\noindent
where

\begin{equation} \label{eq:55}
 t=E-E_{tresh}-i_{\pi}B_{\pi}-j_{\pi}F_{\pi}-i_{\nu}B_{\nu}-
                              j_{\nu}F_{\nu} \: ,
\end{equation}
\begin{equation} \label{eq:56}
 A(p_{\pi},h_{\pi},p_{\nu},h_{\nu},\lambda)=
 \sum_{\lambda_{\nu}=0}^{\lambda} B(p_{\nu},h_{\nu},\lambda_{\nu}) 
                     B(p_{\pi},h_{\pi},\lambda-\lambda_{\nu}) \: ,
\end{equation}
while the coefficients $B(p,h,\lambda)$ are determined by the
Eqs.(\ref{eq:39}-\ref{eq:41}). The total state density obtained as the 
sum of the PSD for all allowed particle-hole numbers is shown in Fig. 
8(a) (Case 17B), as well as the comparison with Kalbach closed 
formula including the above-discussed effective pairing correction, 
the only constant pairing-correction $U_p$ \cite{fu84}, and the 
energy-dependent correction $P_{eff}$, shown in Figs. 8(b) (Cases 
17B/17C) and 8(c) (Cases 17A/17D). Fig. 8(d) (Cases 17A/17D) also 
gives the comparison with the corresponding PSD values obtained 
by means of the Kalbach formula \cite{kalbach8789}, which supports the 
correctness of the approximation for the Pauli-correction term.\\

\noindent
{\it 2.4. Partial state density with surface effects and 
		energy-dependent s.p.l. densities}\\

The surface effects which may be considered within the initial 
target-projectile interaction \cite{kalbach85} by means of the PSD 
formula introduced by Kalbach \cite{kalbach81,kalbach85}

\begin{equation} \label{eq:57}
   \omega(p,h,E,F)=\omega(p,h,E,\infty) f(p,h,E,F) \: ,
\end{equation}
where the finite-depth correction in addition to 
$\omega(p,h,E,\infty)$ given by, e.g.,  Eq. (\ref{eq:27}), is brought 
off by the function

\begin{equation} \label{eq:58}
 f(p,h,E,F)=\sum_{j=0}^h\,(-1)^j\,C_h^j 
     \left({{E-    jF(h)}\over{E}}\right)^{n-1} \Theta(E -jF(h))
\end{equation}
where the Fermi energy $F$=38 MeV corresponds to the central nuclear 
well. The shallower potential in the region of the nuclear surface,
where the first target-projectile interaction in PE reactions is most 
probably localized (see also \cite{gadioli73,avrigeanu96}), is taken 
into account by using an average effective well depth and the related 
Fermi energy $\overline{F}_1$ for the hole number $h\leq 2$

\begin{mathletters}
\begin{eqnarray} \label{eq:59}
 F(h)& = & \overline{F}_1 \hspace*{1.45in}  \mbox{for} \: \: h\leq 2\\
   & = & F_0=38 MeV \hspace*{.71in} \mbox{for} \: \: h>2 \: ,
\end{eqnarray}
\end{mathletters}
so that $\overline{F}_1$-value is taken into account for only the
initial configurations, following the assumption of the surface 
localization of PE two-body interactions exciting them.

   The smaller effective well depth may also determine an increased
significance of another effect, namely the s.p.l. energy-dependence 
of $g(\varepsilon)$. First, by taking into account this dependence,
separate excited-particle state density $g_p$ and single-hole state 
density $g_h$ are involved. Second, the increase in $g_p$ has 
generally compensated for the decrease in $g_h$ except when the 
reduced potential-well depth makes the latter much closer to the
value at the Fermi energy, $g_0=g(F)$; since the excited particles may 
well be excited above the Fermi level in PE reactions, opposite
to the case of the statistical equilibrium, the $g_p$ could increase
significantly. 

The interdependence of the PE surface effects and the energy 
dependence of the s.p.l. density makes the functional form of the 
latter to be yet an open question \cite{kalbach85}. Thus, the 
first-order effects of the FGM dependence have been adopted 
\cite{kalbach85,bogila92,avrigeanu94a} for the time being 

\begin{equation} \label{eq:60}
 g(\varepsilon)=g_0\left({{\varepsilon}\over{F}}\right)^{1/2}=
         {{3A}\over{2F}}\left({{\varepsilon} \over {F}}\right)^{1/2} 
\end{equation}
Next, the actual basic point consists in the use of the average 
excitation energies $\bar{u}_{p}=\bar{\varepsilon}_{p}-F$ for excited 
particles, and $\bar{u}_{h}=F-\bar{\varepsilon}_{h}$ for holes 
\cite{kalbach85}. The former has been estimated to first order from 
the ESM Eq. (\ref{eq:57}), with the result

\begin{equation} \label{eq:61}
 {\overline u_p}={{E}\over{n}}\,{{f(p+1,h,E,F)}\over{f(p,h,E,F)}} \: ,
\end{equation}
while the related one for holes is

\begin{equation} \label{eq:62}
 {\overline u_h} = {{E-p {\overline u_p}}\over{h}} \: ,
\end{equation}
both of them reducing to $E/n$ if the finite depth of the potential 
well is not considered. The corresponding average FGM s.p.l. densities 
for the excited particles and holes 

\begin{mathletters}
\begin{eqnarray} \label{eq:63}
g_p(\bar{u_p})&=&g_0\, \left({{F_0+{\bar u_p}}\over{F_0}}\right)^{1/2}
              = g_p(p,h)\\     
g_h(\bar{u_h})&=&g_0\, \left({{F_0-{\bar u_h}}\over{F_0}}\right)^{1/2}
              = g_h(p,h)     
\end{eqnarray}
\end{mathletters}
are used within the ESM partial state-density (\ref{eq:57}) which 
becomes 

\begin{equation} \label{eq:64}
 \omega(p,h,E,F)={{[g_p(p,h)]^p[g_h(p,h)]^h[E-A_{ph}]^{n-1}}
                 \over{p!h!(n-1)!}} \, f(p,h,E,F) \: .
\end{equation}
It should be underlined that the effective well depth $\overline{F}_1$ 
is involved in evaluating only the quantities $\bar{u}_p$, 
$\bar{u}_h$ and $f(p,h,E,F)$ while the central well depth $F_0$ is 
assumed in calculating the effective s.p.l. densities. The factor 
$g_p^pg_h^h/g_0^n$ shown in Fig. 2 of Ref. \cite{kalbach85} gives just 
the ratio between the PSD including the FGM energy dependence of these 
effective s.p.l. densities, and the ESM state densities with only the 
finite-well depth correction. Both kinds of PLDs are shown in 
Figs. 9(a) (Cases 18/18A) and 9(b) (Cases 18B/18C) for either the 
normal Fermi energy $F_0$ or the average effective value 
$\overline{F}_1$=14 MeV. The analysis concerns the exciton 
configurations corresponding to the first two-body interaction in 
nucleon-induced PE reactions, which are the most important for 
PE-reaction calculations.

The two-component PSD formula taking into account the surface effects 
for the initial target-projectile interaction was given as
\cite{kalbach86} 

\begin{equation} \label{eq:65}
   \omega(p_{\pi},h_{\pi},p_{\nu},h_{\nu},E)=
   {{[g_{\pi}(p)]^{p_{\pi}}[g_{\pi}(h)]^{h_{\pi}} 
     [g_{\nu}(p)]^{p_{\nu}}[g_{\nu}(h)]^{h_{\nu}} 
   [E-A(p_{\pi},h_{\pi},p_{\nu},h_{\nu})]^{n-1}}\over
   {p_{\pi}!h_{\pi}!p_{\nu}!h_{\nu}!(n-1)!}} \, f(p,h,E,F) \: ,
\end{equation}
where 

\begin{equation} \label{eq:66}
  A(p_{\pi},h_{\pi},p_{\nu},h_{\nu})=
   {{p_{\pi m}^2}\over{g_{\pi}}}+
   {{p_{\nu m}^2}\over{g_{\nu}}}-
   {{p_{\pi}^2+h_{\pi}^2+n_{\pi}}\over{4g_{\pi}}}-
   {{p_{\nu}^2+h_{\nu}^2+n_{\nu}}\over{4g_{\nu}}}
\end{equation}
is the simpler form of the Pauli correction for the Fermi level placed 
halfway between the last occupied s.p.l. in the ground state of the 
nucleus and the first vacant s.p.l. The same formalism as in the 
one-fermion system case has been used for the average-excitation 
energies for excited particles and holes. The only additional 
assumption (\ref{eq:46}) concerns the definition of the average 
excitation energies $E_{\pi}$ and $E_{\nu}$, for protons and neutrons 
respectively. 

The PLDs for the two-fermion system exciton configurations 
$(p_{\pi},h_{\pi},p_{\nu},h_{\nu})$=(2100), (1110), (1100), and (1001) 
are also compared in Figs. 9(c) (Cases 18A/19/19A) and 9(d) (Cases 
18C/20/20A) with the one-fermion PSDs of the related configurations 
$(p,h)$=(21) and (11), as well as with the results obtained by using 
the method involving the two-fermion system correction. A behavior 
specific of the particle-hole bound state density arises for the 
two-fermion-system exciton configurations with only one hole, due to 
the use of Eq. (\ref{eq:46}).\\

\noindent
{\it 2.5. Composite (recommended) PSD formulas}\\

The completeness of a composite PSD formula can be obtained by the 
inclusion and optional use of the various ESM corrections :
\begin{itemize} \itemsep 0pt \topsep 0pt \parskip 0pt
\item [-] extension to the case of the finite well depth and bound 
	states \cite{oblozinsky86,mao93,kalbach85};
\item [-] advanced pairing correction by Fu \cite{fu84} and Kalbach
      \cite{kalbach8789} added to the Pauli correction of Williams, 
      proved to be still in good agreement with exact calculations 
      \cite{zhang88,baguer89,mao93};
\item [-] inclusion of the shell effects together with the pairing 
      correction \cite{fu84} and use of the usual level-density 
      parameters also for the PSD/PLD calculation;
\item [-] energy dependence of the single-particle state densities as 
      well as inclusion of the surface effects in the case of the 
      first two steps of multistep processes \cite{kalbach85}, which 
      appear to be most significant for the exciton configurations 
      mainly involved in the description of PE reactions.
\end{itemize}
Therefore, in the one-fermion system case, a composite (recommended) 
PSD formula is 

\begin{equation}\label{eq:2.5.1}
 \omega(p,h,E)={{[g_p(p,h)]^p[g_h(p,h)]^hE^{n-1}}
               \over{p!h!(n-1)!}}f_K(p,h,E,F) \: ,
\end{equation}
where

\begin{mathletters}\label{eq:2.5.2}
\begin{equation}\label{eq:2.5.2a}
 g_p(p,h) = g(F+{\overline u_p})
\end{equation}
\begin{equation}\label{eq:2.5.2b}
 g_h(p,h) = g(F-{\overline u_h}) ,
\end{equation}
\end{mathletters}
with the average excitation energies for particles and holes

\begin{mathletters}\label{eq:2.5.3}
\begin{equation}\label{eq:2.5.3a}
 {\overline u_p} = {{E}\over{n}} {{f_K^+(p,h,E,F)}\over{f_K(p,h,E,F)}}
\end{equation}
\begin{equation}\label{eq:2.5.3b}
 {\overline u_h} ={{E-p \overline u_p}\over{h}} \: ,
\end{equation}
\end{mathletters}
where

\[
 f_K(p,h,E,F) = \sum_{i=0}^p\sum_{j=0}^h (-1)^{i+j}C_p^iC_h^j
              \left({{E-A_K(p,h)-S-iB-jF}\over{E}}\right)^{n-1}
\]
\begin{equation}\label{eq:2.5.4}
    \times \theta(E-E_{thresh}-S-iB-jF) ,
\end{equation}
and

\[
 f_K^+(p,h,E,F)=\sum_{i=0}^p\sum_{j=0}^h (-1)^{i+j}C_p^iC_h^j
       \left( {{E-A_K(p,h)-S-iB-jF}\over{E}} \right)^n 
\]
\begin{equation}\label{eq:2.5.5}
 \times \left(1+{{n}\over{p}} {{iB}\over{E-A_K(p,h)-S-iB-jF}}\right)
  \theta(E-E_{thresh}-S-iB-jF) .
\end{equation}
\bigskip

On the other hand, the two-component partial state density has the 
general expression\\

\[
   \omega(p_{\pi},h_{\pi},p_{\nu},h_{\nu},E)=
 {  {[g_{\pi}(p)]^{p_{\pi}}[g_{\pi}(h)]^{h_{\pi}} 
     [g_{\nu}(p)]^{p_{\nu}}[g_{\nu}(h)]^{h_{\nu}}}\over
    {  p_{\pi}!h_{\pi}!p_{\nu}!h_{\nu}!(n-1)! }    }
 \sum_{i_{\pi}=0}^{p_{\pi}}\, \sum_{j_{\pi}=0}^{h_{\pi}}\,
 \sum_{i_{\nu}=0}^{p_{\nu}}\, \sum_{j_{\nu}=0}^{h_{\nu}}\,
 (-1)^{i_{\pi}+j_{\pi}+i_{\nu}+j_{\nu}}\,
\]
\[
\times
 C_{p_{\pi}}^{i_{\pi}}\, C_{h_{\pi}}^{j_{\pi}}\,
 C_{p_{\nu}}^{i_{\nu}}\, C_{h_{\nu}}^{j_{\nu}}\,
  [E-A_K(p_{\pi},h_{\pi},p_{\nu},h_{\nu})-S-
  i_{\pi}B_{\pi}-j_{\pi}F_{\pi}-i_{\nu}B_{\nu}-j_{\nu}F_{\nu}]^{n-1}
\]
\begin{equation} \label{eq:2.5.6}
  \times  \theta(E-E_{tresh}-S-
  i_{\pi}B_{\pi}-j_{\pi}F_{\pi}-i_{\nu}B_{\nu}-j_{\nu}F_{\nu}) ,
\end{equation}
where

\begin{equation} \label{eq:2.5.7}
  A_K(p_{\pi},h_{\pi},p_{\nu},h_{\nu})=
  A_K(p_{\pi},h_{\pi}) + A_K(p_{\nu},h_{\nu}) , 
\end{equation}
the two-fermion systems being considered at the average excitation 
energies (\ref{eq:46}), respectively. The average s.p.l. densities in 
Eq. (\ref{eq:2.5.6}) are evaluated within the same distinct fermion 
systems. One may observe some difference between the one- and 
two-component formulas concerning the calculation of the nuclear 
potential finite-depth correction, following the original formalisms. 
However, a two-fermion PSD formula quite similar to the one-component 
case and consistent with the average excitation energies (\ref{eq:46}) 
may be obtained with minor changes. 

The {\it optional} provision of the advanced corrections to the ESM 
formulas by Eqs. (\ref{eq:2.5.1}) and (\ref{eq:2.5.6}), respectively, 
is illustrated in Fig. 10. Thus, no correction to the 
Williams formulas for either the one-fermion system, Fig. 10(a) 
(Cases 21/21A), or two-fermion system, Fig. 10(b) (Cases 22/22A), 
determines PLD values obtained with the composite formulas which are 
so close or even identical to the original ones, for the simplest 
configurations $1p1h$. The same is true for the bound-state case with 
only the finite-well depth correction, the comparison with the 
predictions of the Oblo\v{z}insk\'{y} formulas \cite{oblozinsky86} 
being shown for specific configurations as well as the total state 
density given by the corresponding PLDs sum. These results are given 
in Figs. 10(c) (Cases 23/23A) and 10(e) (Cases 23A/24A), respectively, 
for the one-fermion system, and in Figs. 11(a) (Cases 23B/24B/25B) and 
11(c) (Cases 23B/24B) for the two-fermion system formulas.

The specific behavior following the inclusion of the average 
energy-dependent formalism is pointed out by comparison with the 
predictions of the Oblo\v{z}insk\'{y} formulas \cite{oblozinsky86} for 
particular configurations and the total state density given by the 
PLDs sum. This is shown in Figs. 10(d) (Cases 23/23C) and 10(f) (Cases 
23C/24C), respectively, for the one-fermion system, and Figs. 11(b) 
(Cases 23C/24C/25B) and 11(d) (Cases 23D/24D) for the two-fermion 
system formulas. The enhancement of the "bound-state effect" (i.e. 
reaching a PSD maximum value followed by vanishing at a certain 
higher-energy limit) due to the addition of the energy-dependent 
s.p.l. densities is obvious in Figs. 10(d) and 11(b). A particular 
feature is shown comparatively in Figs. 11(a),(b) for the one- and 
two-fermion system configurations with $1p1h$, namely the effects of 
the sequential releasing of the bound-state and finite-well depth 
conditions.

The comparison between the predictions of the composite formula, using 
the FGM energy-dependent s.p.l. densities, and the ESM formula 
including the advanced pairing correction \cite{fu84,kalbach8789} 
is shown in Fig. 12(a) (Cases 26/26A). The same but for the ESM 
one- and two-fermion system formulas including the exact calculation 
of the Pauli correction \cite{zhang88,baguer89,mao93} are shown in 
Figs. 12(c) (Case 28) and 12(d) (Case 28B). The corresponding 
bound-state PLDs are identical as long as the average excitation 
energies of the two kinds of excitons are lower than the 
respective limits, while next the "bound-state effect" is well 
increased within the average energy-dependent formalism.

Finally, the changes due to the inclusion of the surface effects for 
exciton configurations with one and two holes, are illustrated in the 
case of both fermion systems in Figs. 12(a) and Fig. 12(b) (Cases 
27/27B). These changes are higher than, and increasing, the ones due 
to the energy-dependent s.p.l. densities. The comparison with the 
original results \cite{kalbach85} not including the pairing correction
in Fig. 12(b) points out also that taking into account the pairing 
effects is insignificant at higher energies but dominant in the 
threshold region.\\

\noindent
{\it 2.6. The partial level density}\\

\noindent
{\it 2.6.1. One-component Fermi gas formula}\\

   The level density of $p$-particle--$h$-hole configurations of 
excitation energy $E$ and nuclear spin $J$, in the one-fermion 
formulation, is 

\begin{equation} \label{eq:2.6.1}
 \rho(p,h,E,J)=\omega(p,h,E) \, R(n,E,J) \: ,
\end{equation}
where the spin-distribution formula has the general expression
\cite{feshbach80}

\begin{equation} \label{eq:2.6.2}
 R(n,E,J)={{2J+1}\over{2(2\pi)^{1/2}\sigma^3(E,n)}}
       \mbox{exp}\left[{{-(J+1/2)^2}\over{2\sigma^2(E,n)}}\right] \: ,
\end{equation}
and the spin-cutoff factor $\sigma^2(E,n)$ may have the simplified 
energy- and exciton-configuration dependence \cite{fu86}

\begin{equation} \label{eq:2.6.3}
 \sigma^2(E,n)=(\mbox{ln}4)\left({{n}\over{n_c}}\right)
               \left({{E-U_{th}}\over{E}}\right)^x \, \sigma_c^2 \: ,
\end{equation}
where, in addition to the quantities from the previous section,

\begin{equation} \label{eq:2.6.4}
 x=-0.413 + 1.08(n/n_c)^{1/2}-0.226(n/n_c) \: ,
\end{equation}
and $\sigma_c^2$ is the spin-cutoff factor at the critical 
thermodynamic temperature $T_c=2\Delta_0/3.5$, namely 
$\sigma_c^2=gT_c\langle m^2\rangle$. The differences between Eq. 
(\ref{eq:2.6.3}) and the spin-cutoff factor 
$\sigma_F^2(n) \simeq 0.16nA^{2/3}$ initially introduced by Feshbach 
{\it et al.} \cite{feshbach80} concern firstly the threshold existing 
in the former case. Secondly, the result of Feshbach {\it et al.} 
is more appropriate for exciton configurations around the 
most-probable exciton number $\hat n$, which are important to the 
compound-nucleus contribution; at the same time the larger value of 
about 0.28$nA^{2/3}$ obtained by Reffo and Hermann \cite{reffo82} is 
better for the $n$=2 component -- the PE-dominant contribution. 
However, $\sigma^2(E,n)$ can be used for all exciton numbers since 
the respective values are closer to the results of Reffo and Hermann 
around $n$=2, and also in good agreement with $\sigma_F^2(\hat n)$.

  The sum over the nuclear spins, of the PLDs given by Eq. 
(\ref{eq:2.6.1}), corresponds to the partial level density $\rho(p,h,E)$ 
related to the PSD of the same configuration by means of the closed 
formula

\begin{equation} \label{eq:2.6.5}
 \rho(p,h,E)={{\omega(p,h,E)}\over{(2\pi)^{1/2}\sigma(E,n)}} \: .
\end{equation}
On the other hand, the sum of the PLDs over all allowed particle-hole 
numbers $p=h$ should provide the usual nuclear level density involved 
in the statistical model calculations of compound-nucleus processes, 
which has also the one-fermion ESM formula

\begin{equation} \label{eq:2.6.6}
 \rho_1(E,J)=\omega_1(E) \, R(E,J) \: ,
\end{equation}
with the spin-distribution

\begin{equation} \label{eq:2.6.7}
 R(E,J)={{2J+1}\over{2(2\pi)^{1/2}\sigma^3(E)}}
         \mbox{exp}\left[{{-(J+1/2)^2}\over{2\sigma^2(E)}}\right] \: .
\end{equation}
For the spin-cutoff factor $\sigma^2(E)$ in the above equation, which 
should be consistent with the value $\sigma_H^2(E)$ used in 
compound-nucleus (Hauser-Feshbach) calculations, the average of 
$\sigma^2(E,n)$ over $n$ was adopted \cite{fu86} 

\begin{equation} \label{eq:2.6.8}
 \sigma^2(E)={{\sum_{p=h=1} \omega(p,h,E)\,\sigma^2(E,n)}\over
              {\sum_{p=h=1} \omega(p,h,E)}} \: ,
\end{equation}
where the PSD $\omega(p,h,E)$ may be given by, e.g., Eqs. 
(\ref{eq:21}) or (\ref{eq:28}). This average was also proved to be
consistent with the general form of $\sigma_H^2(E)$ \cite{dilg73}

\begin{equation} \label{eq:2.6.9}
 \sigma_H^2(E)=g \langle m^2 \rangle t \: ,
\end{equation}
where the value $\langle m^2 \rangle$=0.24$A^{2/3}$ was assumed
\cite{fu86}.

  Furthermore, the sum of the PLDs over both $p$ (with the restriction 
$p$=$h$) and spins gives the total level density 

\begin{equation} \label{eq:2.6.10}
 \rho_1(E)=\sum_{p=h=1} \sum_J \rho(p,h,E,J)
\end{equation}
which is related to the one-fermion ESM total state density by

\begin{equation} \label{eq:2.6.11}
 \rho_1(E)={{\omega_1(E)}\over{(2\pi)^{1/2}\sigma(E)}} \: .
\end{equation}
The comparison of the values given by the above closed formula and 
the sum (\ref{eq:2.6.10}) (i.e. Eqs. (52) and (53) respectively of Ref.
\cite{fu86}) for the excited nucleus $^{41}$Ca is shown in Fig. 13(a) 
(Case 11).

The partial and total state and level densities corresponding to the
two-component ESM formulas can be obtained by using either the  
renormalization method \cite{fu86,akkermans85} or the adjustment 
method of the two-fermion formula parameters \cite{fu84,fu86}. The
former, illustrated in Fig. 4(d), offers the advantage of using only 
one set of e.g. BSFG model parameters for both the PLDs involved in 
PE calculations and the nuclear level densities for Hauser-Feshbach 
calculations.\\

\noindent
{\it 2.6.2. Two-component Fermi gas formula}\\

The two-fermion level density formula is \cite{fu91}

\begin{equation} \label{eq:2.6.12}
\rho(p_{\pi},h_{\pi},p_{\nu},h_{\nu},E,J)=
\omega(p_{\pi},h_{\pi},p_{\nu},h_{\nu},E,J) \, R(n_{\pi},n_{\nu},E,J)\:,
\end{equation}
where the spin-distribution formula has the similar general expression
\cite{feshbach80}

\begin{equation} \label{eq:2.6.13}
 R(n_{\pi},n_{\nu},E,J)=
     {{2J+1}\over{2(2\pi)^{1/2}\sigma^3(E,n_{\pi},n_{\nu})}}
     \mbox{exp}\left[{{-(J+1/2)^2}\over{2\sigma^2(E,n_{\pi},n_{\nu})}}
     \right] \: ,
\end{equation}
and the spin-cutoff factor $\sigma^2(E,n_{\pi},n_{\nu})$ for two kinds
of fermions is defined as the sum of the two one-fermion components

\begin{equation} \label{eq:2.6.14}
  \sigma^2(E,n_{\pi},n_{\nu})=
  \sigma^2(E_{\pi},n_{\pi})+\sigma^2(E_{\nu},n_{\nu}) \: .
\end{equation}
The mean-gap approximation and the approximations (\ref{eq:46}) for 
$E_{\pi}$ and $E_{\nu}$ as well as the parameterized function for the 
one-fermion system \cite{fu86} are used in this respect.

The sum of the two-fermion PLDs over spins and both $p_{\pi}$ and 
$p_{\nu}$ (with the restriction $p_{\pi}=h_{\pi}$ and 
$p_{\nu}=h_{\nu}$) provides the total level density 

\begin{equation} \label{eq:2.6.15}
 \rho_2(E)=\sum_{p_{\pi}=h_{\pi},p_{\nu}=h_{\nu}} \sum_J 
 \rho(p_{\pi},h_{\pi},p_{\nu},h_{\nu},E,J) \: 
\end{equation}
which is related to the two-fermion ESM total state density by

\begin{equation} \label{eq:2.6.16}
 \rho_2(E)={{\omega_2(E)}\over{(2\pi)^{1/2}\sigma(E)}} \: .
\end{equation}
The same averages of $\sigma^2(E,n)$ over $n$ as for the one-fermion 
formulas \cite{fu86} are used for the above spin-cutoff factor 

\begin{equation} \label{eq:2.6.17}
 \sigma^2(E)={{\sum_{p_{\pi}=h_{\pi},p_{\nu}=h_{\nu}} 
 \omega(p_{\pi},h_{\pi},p_{\nu},h_{\nu},E)
 \,\sigma^2(E,n)}\over{\sum_{p_{\pi}=h_{\pi},p_{\nu}=h_{\nu}} 
 \omega(p_{\pi},h_{\pi},p_{\nu},h_{\nu},E)}} \: ,
\end{equation}
where $\omega(p_{\pi},h_{\pi},p_{\nu},h_{\nu},E)$ is given by Eq. 
(\ref{eq:43}). 

The comparison of the values given by the above closed formula and the 
sum (\ref{eq:2.6.15}) for the excited nucleus $^{41}$Ca is shown in 
Fig. 13(b) (Case 11B). Fig. 14 (Cases 11/11B) shows the comparison 
between the average spin-cutoff values 
$\sigma_2^2(n$=$2)$=$[\sigma_2^2(E,n_{\pi}$=$2,n_{\nu}$=$0) +
                      \sigma_2^2(E,n_{\pi}$=$0,n_{\nu}$=$2)]/2$
\cite{fu91} and $\sigma_1^2(n=2)$ corresponding to the one-fermion 
system, as well as of the average values (\ref{eq:2.6.17}) noted 
$\sigma_2^2$ and the corresponding $\sigma_1^2$ (both of them noted 
"all n" in Figs. 13 and 14). The related PLD and total level density 
values are shown in Fig. 13(c) (Cases 11/11B) while the comparison 
between the two-fermion system PLD values and the ones obtained by 
using the two-fermion correction \cite{fu86,akkermans85} is given in 
Fig. 13(d) (Cases 11B/11E).\\

\noindent
{\bf 3. Program organization}\\

The program PLD.FOR is the collection of the above-described 
algorithms developed until now and widely used for PSD/PLD within 
nuclear model calculations. It also includes the recommended 
(combined) PLD formulas given in this work. Fourteen different PSD 
formulas are available as FORTRAN77 functions to be {\it used as they 
are} in various applications or codes. The one method involved in
calculating the PLD is given within a subroutine only for the tabular 
printing of the results, while the replacement of the SUBROUTINE 
statement by the FUNCTION one is immediate.

Second, the PLD.FOR has been organized so that various formulas and 
versions may be tried as well as the comparison between their 
predictions. This could also be useful for further development of the 
PSD calculation methods. 

Since the first aim of this work has been to provide tools for users 
of PSD/PLD, the optimization of the respective procedures is made 
firstly in this respect. The possibility of using these functions 
independently has the related drawback of increasing the execution 
time. A proper use, e.g., of the PLD.FOR for PSD calculation with 
exact Pauli-correction term of Mao Ming De and Guo Hua \cite{mao93} 
would involve the calculation of the exact coefficients 
$B(p,h,\lambda)$ only once in the main program. One should keep this 
aspect in mind if PLD.FOR will be used on low-speed PCs.\\

\noindent
{\it 3.1. Subprograms}\\

The MAIN program reads the input data which are listed and described 
in Table 1 in reading order (including the names of variables which 
are also used below and stand for various quantities given in the 
previous section). The formatted read is used for all data just 
to make possible the input of only few of them, while by-default 
values may be involved for the rest. Then, the partial state densities 
$w(p,h,E)$ are calculated by using the specified formula, as well as
eventually the related partial level densities $D(p,h,E,J)$. In the 
general case when the calculation is carried on for all pairs of the
exciton numbers $p$=$h$, it is also done for (i) the nuclear state 
density $w(E)$, as the PSD-sum over all allowed exciton numbers for 
which the PSD is higher than the value WMINACC=0.1 MeV$^{-1}$, or
(ii) the total level density $D(E)$ as the PLD-sum over the exciton
numbers and the nuclear angular momentum $J$. At the same time the 
corresponding values $Wasym(E)$ or $Dasym(E)$, respectively, are 
calculated by means of the ESM closed formulas, in order to make 
possible a test of the overall consistency. The two-fermion system 
correction \cite{akkermans85} is involved optionally when the 
one-component Fermi gas formulas are used. 

The subroutine {\bf PRINTIN} prints the type of the PSD/PLD formula 
and the parameters used in calculation.
 
The subroutine {\bf PRINTWN} tabulates the calculated values 
$w(p,h,E)$ or $D(p,h,E)$ (the latter being the sum over $J$ of 
$D(p,h,E,J)$, i.e. the total level density for a given exciton 
configuration), for either (i) some given particle-hole 
configurations, or (ii) all pairs of equal numbers of excited 
particles and holes. In the latter case the values $w(E)$ or $D(E)$ 
are also printed within a first table including the PSD/PLDs
for $p$=$h$ from 1 to 7, while the corresponding closed-formula values 
$Wasym(E)$ or $Dasym(E)$ are given within the second table including 
the PSD/PLDs for $p$=$h$ from 8 to 16. A table of the values 
$D(p,h,E,J)$, but only for the last excitation energy $E$ involved in 
one calculation, is also printed in the case of the PLD calculation 
for a particular exciton configuration.

The functions {\bf WIL1} and {\bf WIL2} calculate the partial state 
density $w(p,h,E)$ for a given exciton configuration, by using the 
Williams \cite{williams71} one- and two-fermion system formulas, 
respectively.

The functions {\bf WOB1} and {\bf WOB2} calculate the partial state 
density for a given particle-hole configuration by means of the 
B\v{e}t\'{a}k-Dobe\v{s} one- and two-fermion system formulas, 
respectively \cite{betak76}, with the nuclear potential finite-depth 
correction. The calculation of the bound-state density according to 
the Oblo\v{z}insk\'{y} formulas \cite{oblozinsky86} is carried out if 
a value is specified for the nucleon binding energy.

The functions {\bf WFU1} and {\bf WFU2} calculate the PSD for a 
given exciton configuration by means of the one- and two-fermion 
system formulas, respectively, including the advanced pairing 
correction by Fu \cite{fu84}.

The functions {\bf WK1} and {\bf WK2} calculate the PSD for a given 
exciton configuration by using the improved implementation of the 
pairing correction by Kalbach \cite{kalbach8789} within the one- and 
two-fermion system formulas, respectively.

The functions {\bf WK3} and {\bf WK4} calculate the PSD for a given 
exciton configuration by using the FGM energy-dependence of the 
single-excited particle and single-hole state densities, and/or the 
finite-depth correction including the nuclear-surface effects 
introduced by Kalbach \cite{kalbach85} within the one- and two-fermion 
system formulas, respectively.

The functions {\bf WM1} and {\bf WM2} calculate the PSD for a given 
exciton configuration by using the exact calculation of the 
Pauli-exclusion effect \cite{mao93} and the pairing correction by 
Kalbach \cite{kalbach8789} within the one- and two-fermion system 
formulas, respectively.

The functions {\bf WR1} and {\bf WR2} calculate the PSD for a given 
exciton configuration by using the composite (recommended) formalism
including optionally (i) the improved implementation of the pairing 
correction by Kalbach \cite{kalbach8789}, (ii) the FGM 
energy-dependence of the single-excited particle and single-hole state 
densities, and/or (iii) the finite-depth correction including the 
nuclear-surface effects introduced by Kalbach \cite{kalbach85}, within
the one- and two-fermion system formulas, respectively.

The function {\bf PFU} calculates the advanced pairing correction by 
Fu \cite{fu84}, for the one-component Fermi-gas.

The function {\bf AK} calculates the advanced pairing correction by 
Kalbach \cite{kalbach8789}, for the one-component Fermi-gas.

The function {\bf FDC0} calculates the nuclear potential 
finite-depth correction factor $f(p,h,E,F)$ \cite{kalbach85} for the 
one-component Fermi-gas.

The function {\bf FDC} calculates the nuclear potential finite-depth 
correction factor $f(p+1,h,E,F)$ \cite{kalbach85} but for the case of 
bound states.

The subroutine {\bf SUBPLD} calculates the partial level density 
$D(p,h,E,J)$ for a given particle-hole configuration, excitation 
energy $E$ and nuclear spin $J$, as well as the respective total level
density $D(p,h,E)$ as their sum over $J$, by using the partial state 
density $w(p,h,E)$ and the formalism of Fu \cite{fu86,fu91}.

The function {\bf SIG2FU} calculates the spin-cutoff factor for a 
given excited particle-hole configuration \cite{fu86}.

The function {\bf FCTR} calculates the factorial of natural 
numbers.\\

\noindent
{\bf 4. An illustrative test run}\\

The sample case 23C (see also Fig. 10) is here discussed since it 
documents few of the specific features of the program PLD at once. 
Thus, the particle-hole bound state densities for some of the 
few-exciton configurations analyzed by Oblo\v{z}insk\'{y} (Fig. 1 of 
\cite{oblozinsky86}) are calculated by  using both the composite 
formula and the one of Oblo\v{z}insk\'{y}. In the former case the FGM 
energy dependence for the s.p.l. density is additionally taken into 
account, while the input data correspond to the reference work 
\cite{oblozinsky86}. The comparison of the results obtained with the
two formulas is possible within the same table with the columns of 
results given in the input-data reading order (the table in the 
attached output copy below is reduced to the first half of the 
involved excitation energies). The corresponding curves are shown in 
Fig. 10(d). One note should concern the printed type of the PSD 
formula used: it denotes only the last one when the option parameter 
ICONT=2 is used and more than one formula are involved in calculation.

An additional calculation is included in this case with respect to 
its title. To make possible the comparison between the PSD for a given
exciton configuration with equal numbers of holes and excited 
particles which is calculated both specifically and within the general
case for all pairs, the latter calculation is also made. Moreover, in
the attached copy of the reduced output only the first two tables for 
this calculation are shown (which would correspond to the use of the 
option parameter value ICONT$=$-1). One may thus find the printed 
values of the total state density $w(E)$ obtained as the PSD-sum, in 
the first table, and the related $Wasym(E)$ closed-formula values, 
in the second table. The corresponding curves are shown in Fig. 10(e).
However, this part of the case output consists of four tables which 
correspond to the maximum numbers $p$=$h$=25 for which the PSD value 
at higher limit of excitation energies is yet higher than WMINACC. 

Actually, the PSD/PLD values for exciton numbers higher than $\hat n$ 
are usually of less interest, while the PLD-program output includes 
them just for the sake of completeness. In the possible case that an 
output table would include only zero values, it is omitted from print.
We may add that questions may arise for calculations for all pairs 
$p$=$h$ when quite large excitation energies and the two-fermion 
system formulas are involved, due to the limits of vectors. However, 
the use of the respective FUNCTION subprograms for given exciton 
numbers and energy, i.e. the usual case of nuclear reaction 
cross-section calculations, is always straightforward.\\

\noindent
{\bf Acknowledgements}\\

The authors are grateful to Pavel Oblo\v{z}insk\'{y} and Mark Chadwick
who have encouraged this work. Assistance of Arpad Harangozo with the 
functions including the exact correction for the Pauli-exclusion 
principle is acknowledged. This work has been carried out under the 
Research Contracts No. 8886/R0-R1/RBF of the International Atomic 
Energy Agency (Vienna) and the Romanian Ministry of Research and 
Technology Contract No. 4/A12.

\newpage

{\bf TEST RUN (REDUCED) OUTPUT}

{\tighten
\begin{verbatim}
 PARTIAL STATE/LEVEL DENSITIES CALCULATED FROM THE FOLLOWING PARAMETERS
 **********************************************************************

 Case 23C: C-formula(g-FGM)/Oblozinsky(1986): one-f. bound PSD for configs.

  NUCLEUS:  CHARGE NO. Z= 0.  MASS NO. A=  0.  PAIRING Up=  .000 MEV

  ONE-FERMION FORMULA:
    P.OBLOZINSKY, NUCL.PHYS. A453,127(1986), Eqs.(7,9)

  SINGLE-PARTICLE STATE DENSITY: G= 8.000 /MEV
  FERMI ENERGY: F=  32.000 MEV
  NUCLEON BINDING ENERGY: B=   8.000 MEV
 ______________________________________________________________________________
 ENERGY                              w(p,h,E)
  (MeV)                               (1/MeV)
 ______________________________________________________________________________
 (p,h)= 1  1     0  2     2  1     1  2     1  1     0  2     2  1     1  2
 ______________________________________________________________________________
   1.00 64.0     21.2     70.8     70.1     64.0     30.0     113.     113.    
   2.00 128.     51.9     393.     385.     128.     62.0     481.     481.    
   3.00 192.     81.6     976.     947.     192.     94.0     .110E+04 .110E+04
   4.00 255.     110.     .183E+04 .175E+04 256.     126.     .198E+04 .198E+04
   5.00 319.     138.     .294E+04 .280E+04 320.     158.     .312E+04 .312E+04
   6.00 382.     164.     .433E+04 .407E+04 384.     190.     .451E+04 .451E+04
   7.00 445.     190.     .600E+04 .558E+04 448.     222.     .616E+04 .616E+04
   8.00 508.     215.     .794E+04 .731E+04 512.     254.     .806E+04 .806E+04
   9.00 499.     238.     .999E+04 .917E+04 512.     286.     .100E+05 .101E+05
  10.00 490.     261.     .118E+05 .110E+05 512.     318.     .117E+05 .122E+05
  11.00 480.     282.     .132E+05 .127E+05 512.     350.     .131E+05 .142E+05
  12.00 470.     303.     .144E+05 .144E+05 512.     382.     .143E+05 .163E+05
  13.00 460.     323.     .153E+05 .160E+05 512.     414.     .152E+05 .183E+05
  14.00 450.     341.     .158E+05 .175E+05 512.     446.     .158E+05 .204E+05
  15.00 440.     359.     .160E+05 .190E+05 512.     478.     .162E+05 .224E+05
  16.00 429.     375.     .160E+05 .204E+05 512.     510.     .164E+05 .244E+05
  17.00 418.     391.     .156E+05 .217E+05 512.     542.     .164E+05 .265E+05
  18.00 407.     405.     .153E+05 .230E+05 512.     574.     .164E+05 .285E+05
  19.00 396.     419.     .149E+05 .242E+05 512.     606.     .164E+05 .306E+05
  20.00 384.     432.     .146E+05 .253E+05 512.     638.     .164E+05 .326E+05
  21.00 372.     443.     .142E+05 .264E+05 512.     670.     .164E+05 .347E+05
  22.00 359.     454.     .138E+05 .274E+05 512.     702.     .164E+05 .367E+05
  23.00 346.     463.     .134E+05 .283E+05 512.     734.     .164E+05 .388E+05
  24.00 333.     472.     .130E+05 .292E+05 512.     766.     .164E+05 .408E+05
  25.00 318.     480.     .126E+05 .300E+05 512.     798.     .164E+05 .429E+05
  26.00 304.     486.     .122E+05 .307E+05 512.     830.     .164E+05 .449E+05
  27.00 288.     492.     .117E+05 .313E+05 512.     862.     .164E+05 .470E+05
  28.00 272.     496.     .113E+05 .319E+05 512.     894.     .164E+05 .490E+05
  29.00 254.     500.     .108E+05 .324E+05 512.     926.     .164E+05 .511E+05
  30.00 235.     503.     .103E+05 .329E+05 512.     958.     .164E+05 .531E+05
  31.00 215.     504.     .978E+04 .333E+05 512.     990.     .164E+05 .552E+05
  32.00 192.     505.     .922E+04 .336E+05 512.     .102E+04 .164E+05 .572E+05
  33.00 158.     487.     .861E+04 .338E+05 448.     994.     .163E+05 .590E+05
  34.00 126.     457.     .793E+04 .337E+05 384.     962.     .159E+05 .604E+05
  35.00 96.8     427.     .718E+04 .332E+05 320.     930.     .153E+05 .612E+05
  36.00 69.7     399.     .637E+04 .325E+05 256.     898.     .144E+05 .614E+05
  37.00 45.6     371.     .551E+04 .316E+05 192.     866.     .133E+05 .612E+05
  38.00 25.0     344.     .463E+04 .303E+05 128.     834.     .119E+05 .605E+05
  39.00 8.89     319.     .375E+04 .289E+05 64.0     802.     .102E+05 .592E+05
  40.00 .000     294.     .291E+04 .272E+05 .000     770.     .832E+04 .575E+05
                                                         1998- 1-10 10:15:42:99
                                                         EXEC TIME=    .2 S

 PARTIAL STATE/LEVEL DENSITIES CALCULATED FROM THE FOLLOWING PARAMETERS
 **********************************************************************

 Case 23C: C-formula(g-FGM)/Oblozinsky(1986): one-f. bound PSD for configs.

  NUCLEUS:  CHARGE NO. Z= 0.  MASS NO. A=  0.  PAIRING Up=  .000 MEV

  ONE-FERMION FORMULA:
    COMPOSITE FORMULA

  SINGLE-PARTICLE STATE DENSITY: G= 8.000 /MEV
  FERMI ENERGY: F=  32.000 MEV
  AV. EFF. FERMI ENERGY/1ST-2ND STEPS: F1=  32.000MEV
  NUCLEON BINDING ENERGY: B=   8.000 MEV
 ______________________________________________________________________________
 ENERGY   w(E)                              w(p,h,E)
  (MeV) (1/MeV)                              (1/MeV)
                _______________________________________________________________
                p=h=1        2        3        4        5        6        7
 ______________________________________________________________________________
   1.00 173.     64.0     109.     .000     .000     .000     .000     .000    
   2.00 .183E+04 128.     .111E+04 589.     .000     .000     .000     .000    
   3.00 .124E+05 192.     .402E+04 .707E+04 .116E+04 .000     .000     .000    
   4.00 .657E+05 255.     .987E+04 .365E+05 .183E+05 818.     .000     .000    
   5.00 .293E+06 319.     .197E+05 .125E+06 .128E+06 .194E+05 170.     .000    
   6.00 .115E+07 382.     .344E+05 .334E+06 .584E+06 .191E+06 .775E+04 .000    
   7.00 .411E+07 445.     .551E+05 .760E+06 .202E+07 .116E+07 .118E+06 900.    
   8.00 .136E+08 508.     .828E+05 .154E+07 .578E+07 .514E+07 .995E+06 .265E+05
   9.00 .419E+08 499.     .118E+06 .285E+07 .144E+08 .184E+08 .582E+07 .350E+06
  10.00 .122E+09 490.     .160E+06 .493E+07 .322E+08 .559E+08 .264E+08 .287E+07
  11.00 .341E+09 480.     .208E+06 .807E+07 .663E+08 .150E+09 .989E+08 .172E+08
  12.00 .912E+09 470.     .260E+06 .126E+08 .127E+09 .365E+09 .320E+09 .815E+08
  13.00 .235E+10 460.     .316E+06 .187E+08 .231E+09 .819E+09 .924E+09 .325E+09
  14.00 .585E+10 450.     .373E+06 .269E+08 .399E+09 .172E+10 .242E+10 .113E+10
  15.00 .142E+11 440.     .430E+06 .373E+08 .660E+09 .340E+10 .588E+10 .349E+10
  16.00 .333E+11 429.     .486E+06 .503E+08 .105E+10 .640E+10 .133E+11 .986E+10
  17.00 .765E+11 418.     .541E+06 .660E+08 .161E+10 .115E+11 .285E+11 .257E+11
  18.00 .172E+12 407.     .593E+06 .845E+08 .240E+10 .200E+11 .578E+11 .627E+11
  19.00 .377E+12 396.     .643E+06 .106E+09 .348E+10 .334E+11 .112E+12 .144E+12
  20.00 .813E+12 384.     .690E+06 .130E+09 .492E+10 .540E+11 .210E+12 .315E+12
  21.00 .172E+13 372.     .735E+06 .157E+09 .678E+10 .849E+11 .378E+12 .656E+12
  22.00 .358E+13 359.     .778E+06 .187E+09 .915E+10 .130E+12 .658E+12 .131E+13
  23.00 .733E+13 346.     .819E+06 .219E+09 .121E+11 .194E+12 .111E+13 .253E+13
  24.00 .148E+14 333.     .857E+06 .254E+09 .157E+11 .284E+12 .183E+13 .472E+13
  25.00 .294E+14 318.     .893E+06 .291E+09 .201E+11 .406E+12 .293E+13 .853E+13
  26.00 .577E+14 304.     .927E+06 .329E+09 .253E+11 .569E+12 .459E+13 .150E+14
  27.00 .112E+15 288.     .958E+06 .370E+09 .314E+11 .784E+12 .703E+13 .256E+14
  28.00 .214E+15 272.     .987E+06 .412E+09 .384E+11 .106E+13 .106E+14 .428E+14
  29.00 .406E+15 254.     .101E+07 .456E+09 .465E+11 .142E+13 .155E+14 .698E+14
  30.00 .761E+15 235.     .104E+07 .501E+09 .557E+11 .186E+13 .225E+14 .111E+15
  31.00 .141E+16 215.     .106E+07 .548E+09 .659E+11 .241E+13 .320E+14 .174E+15
  32.00 .259E+16 192.     .108E+07 .595E+09 .774E+11 .308E+13 .447E+14 .268E+15
  33.00 .471E+16 158.     .110E+07 .643E+09 .900E+11 .389E+13 .617E+14 .404E+15
  34.00 .849E+16 126.     .111E+07 .692E+09 .104E+12 .487E+13 .839E+14 .600E+15
  35.00 .152E+17 96.8     .112E+07 .741E+09 .119E+12 .602E+13 .113E+15 .877E+15
  36.00 .269E+17 69.7     .113E+07 .791E+09 .135E+12 .737E+13 .149E+15 .126E+16
  37.00 .473E+17 45.6     .113E+07 .840E+09 .153E+12 .894E+13 .196E+15 .179E+16
  38.00 .825E+17 25.0     .112E+07 .890E+09 .172E+12 .108E+14 .253E+15 .251E+16
  39.00 .143E+18 8.89     .110E+07 .939E+09 .192E+12 .128E+14 .325E+15 .347E+16
  40.00 .245E+18 .000     .108E+07 .987E+09 .214E+12 .152E+14 .412E+15 .474E+16
 ______________________________________________________________________________
 ENERGY Wasym(E)                            w(p,h,E)
  (MeV) (1/MeV)                              (1/MeV)
                _______________________________________________________________
                p=h=8        9       10       11       12       13       14
 ______________________________________________________________________________
   1.00 204.     .000     .000     .000     .000     .000     .000     .000    
   2.00 .206E+04 .000     .000     .000     .000     .000     .000     .000    
   3.00 .138E+05 .000     .000     .000     .000     .000     .000     .000    
   4.00 .723E+05 .000     .000     .000     .000     .000     .000     .000    
   5.00 .321E+06 .000     .000     .000     .000     .000     .000     .000    
   6.00 .126E+07 .000     .000     .000     .000     .000     .000     .000    
   7.00 .447E+07 .000     .000     .000     .000     .000     .000     .000    
   8.00 .147E+08 .000     .000     .000     .000     .000     .000     .000    
   9.00 .455E+08 .153E+04 .000     .000     .000     .000     .000     .000    
  10.00 .133E+09 .415E+05 .000     .000     .000     .000     .000     .000    
  11.00 .370E+09 .552E+06 993.     .000     .000     .000     .000     .000    
  12.00 .989E+09 .474E+07 .308E+05 .000     .000     .000     .000     .000    
  13.00 .255E+10 .302E+08 .462E+06 219.     .000     .000     .000     .000    
  14.00 .635E+10 .154E+09 .444E+07 .100E+05 .000     .000     .000     .000    
  15.00 .154E+11 .660E+09 .313E+08 .194E+06 .000     .000     .000     .000    
  16.00 .362E+11 .247E+10 .176E+09 .225E+07 .112E+04 .000     .000     .000    
  17.00 .833E+11 .826E+10 .827E+09 .185E+08 .348E+05 .000     .000     .000    
  18.00 .187E+12 .252E+11 .338E+10 .118E+09 .554E+06 .000     .000     .000    
  19.00 .412E+12 .707E+11 .123E+11 .628E+09 .577E+07 .191E+04 .000     .000    
  20.00 .890E+12 .186E+12 .407E+11 .286E+10 .446E+08 .538E+05 .000     .000    
  21.00 .189E+13 .458E+12 .124E+12 .115E+11 .276E+09 .822E+06 .000     .000    
  22.00 .394E+13 .107E+13 .351E+12 .417E+11 .144E+10 .843E+07 .126E+04 .000    
  23.00 .811E+13 .240E+13 .933E+12 .138E+12 .654E+10 .652E+08 .390E+05 .000    
  24.00 .164E+14 .514E+13 .235E+13 .426E+12 .265E+11 .408E+09 .644E+06 .000    
  25.00 .328E+14 .106E+14 .562E+13 .122E+13 .971E+11 .216E+10 .705E+07 287.    
  26.00 .647E+14 .210E+14 .129E+14 .332E+13 .328E+12 .100E+11 .577E+08 .124E+05
  27.00 .126E+15 .405E+14 .283E+14 .855E+13 .103E+13 .416E+11 .380E+09 .253E+06
  28.00 .243E+15 .757E+14 .601E+14 .210E+14 .303E+13 .157E+12 .211E+10 .324E+07
  29.00 .462E+15 .138E+15 .123E+15 .494E+14 .845E+13 .545E+12 .102E+11 .299E+08
  30.00 .872E+15 .244E+15 .245E+15 .112E+15 .224E+14 .176E+13 .443E+11 .217E+09
  31.00 .163E+16 .423E+15 .473E+15 .244E+15 .565E+14 .536E+13 .174E+12 .131E+10
  32.00 .301E+16 .716E+15 .889E+15 .517E+15 .137E+15 .154E+14 .631E+12 .683E+10
  33.00 .551E+16 .119E+16 .163E+16 .106E+16 .320E+15 .421E+14 .213E+13 .315E+11
  34.00 .100E+17 .193E+16 .292E+16 .211E+16 .721E+15 .110E+15 .672E+13 .132E+12
  35.00 .180E+17 .308E+16 .512E+16 .411E+16 .157E+16 .276E+15 .201E+14 .504E+12
  36.00 .322E+17 .483E+16 .879E+16 .778E+16 .333E+16 .665E+15 .571E+14 .179E+13
  37.00 .572E+17 .745E+16 .148E+17 .144E+17 .687E+16 .155E+16 .155E+15 .594E+13
  38.00 .101E+18 .113E+17 .245E+17 .261E+17 .138E+17 .350E+16 .403E+15 .186E+14
  39.00 .176E+18 .169E+17 .398E+17 .465E+17 .270E+17 .767E+16 .101E+16 .553E+14
  40.00 .306E+18 .250E+17 .637E+17 .809E+17 .517E+17 .163E+17 .244E+16 .157E+15
 ______________________________________________________________________________
                                                         1998- 1-10 10:15:44:20
                                                         EXEC TIME=   1.2 S
\end{verbatim}  }

\newpage
\noindent
{\bf Figure Captions}

\begin{itemize} \itemsep 0pt \topsep 0pt \parskip 0pt
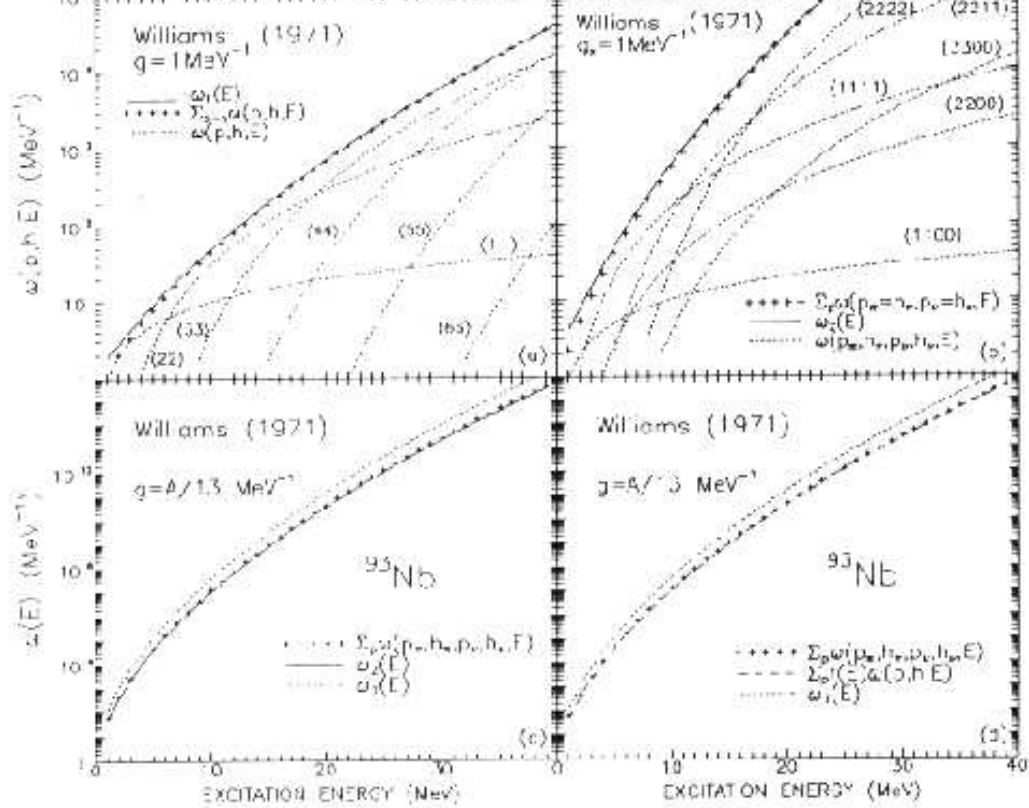
\item [FIG. 1.] Particle-hole state densities for the given $ph$ 
configurations, and the sum of the PSD for all allowed 
exciton numbers $p$=$h$ (crosses), obtained with (a) the Williams' 
one- and (b)-(d) two-fermion formulas for (a) the generic value $g$=1 
MeV$^{-1}$, (b) $g_{\pi}$=$g_{\nu}$=$g$/2=1 MeV$^{-1}$, and (c),(d) 
the nucleus $^{93}$Nb and the phenomenological value $g$=$A/13$ 
MeV$^{-1}$. There are also shown the total nuclear state-densities 
given by the ESM formulas for (a) the one- and (b)-(d) two-fermion 
systems \cite{ericson60} (solid curves), compared with (c),(d) the 
one-fermion system closed formula (dotted curves) and (d) the sum of 
the renormalized one-fermion PSDs \cite{akkermans85} (dashed curve).

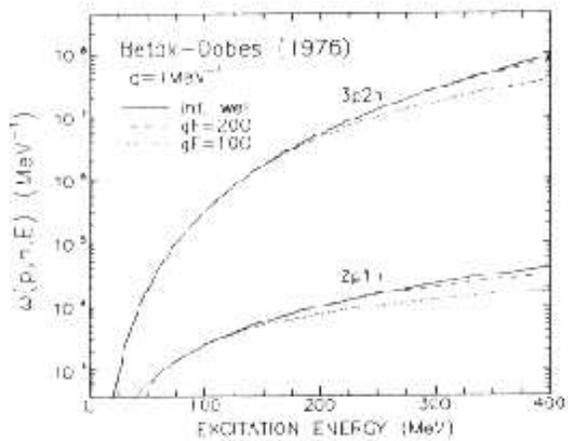
\item [FIG. 2.] Particle-hole state densities for the $2p1h$ and 
$3p2h$ configurations, obtained with the ESM one-fermion formula with
(dashed and dotted curves, for various $F$ values) and without (solid 
curve) the potential finite-depth correction \cite{betak76}, for the 
value $g$=1 MeV$^{-1}$.

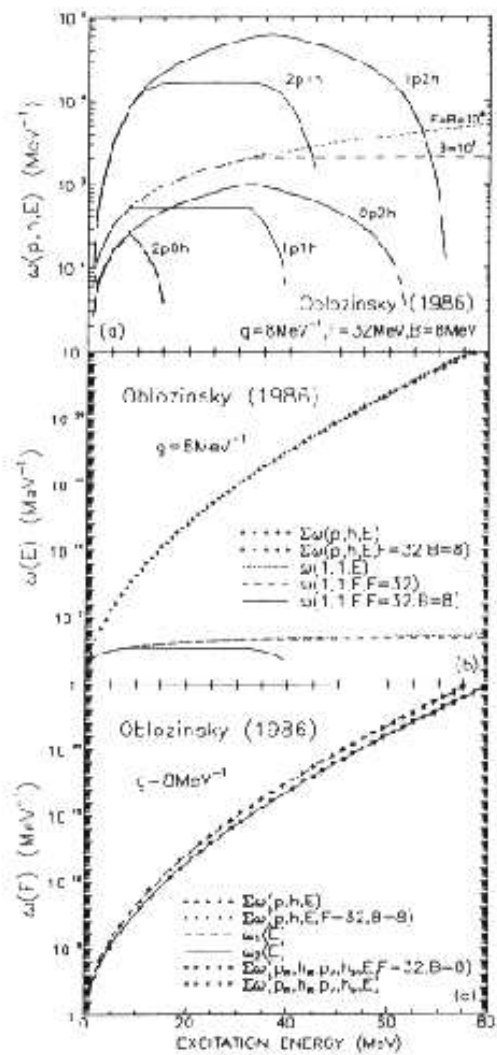
\item [FIG. 3.] Particle-hole bound state densities obtained with 
the Williams' one-fermion formula including the nuclear-potential 
finite depth \cite{oblozinsky86} for (a) the given $ph$ configurations  
(solid curves), and (a),(b) the configuration $1p1h$, releasing 
consecutively the finite-depth potential (dashed curve) and the bound 
state conditions (dotted curve) by means of large $F$- and 
$B$-values, (b),(c) their sum for all allowed exciton numbers $p$=$h$ 
without/with the two corrections (upper/lower crosses), and (c)
the similar sum of the two-fermion PSDs (upper/lower x) compared 
with the one- and two-fermion system closed formulas (dashed and solid 
curves, respectively), for the values $g$=8 MeV$^{-1}$, $F$=32 MeV, 
and $B$= 8 MeV.

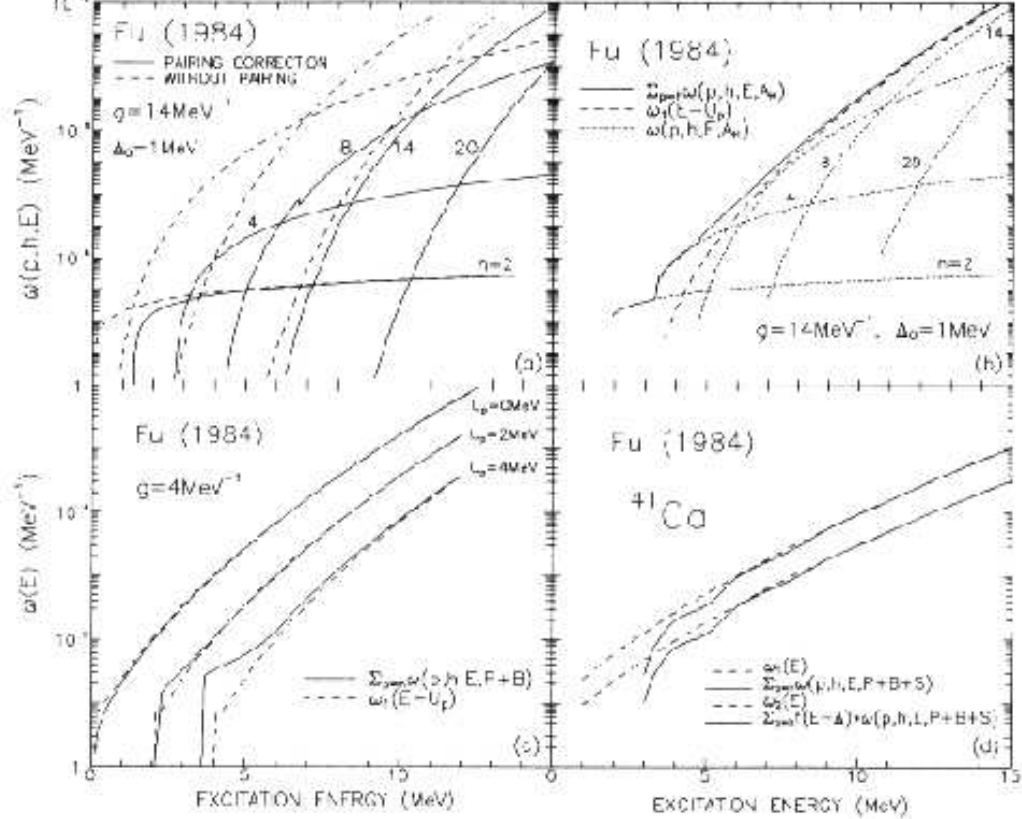
\item [FIG. 4.] PSDs obtained with the Williams one-fermion 
formula (a) without/with the advanced pairing correction \cite{fu84} 
(dashed/solid curves) and (b) above the threshold energy for each 
configuration (dotted curves), leading to the sum of the PSD for all 
allowed exciton numbers $p$=$h$ (solid curve), compared with the 
closed formula (dashed curve) for the values $g$=14 MeV$^{-1}$ and 
$\Delta_0$=1 MeV, as well as (c) for $g$=4 MeV$^{-1}$ and constant 
pairing correction $U_p$ of 0, 2, and 4 MeV, and (d) for the 
one/two-fermion system formulas (upper/lower respective curves) and 
the BSFG parameters $a$=4.12 MeV$^{-1}$ and $\Delta$=-1.07 MeV for the 
excited nucleus $^{41}$Ca \cite{fu84}.

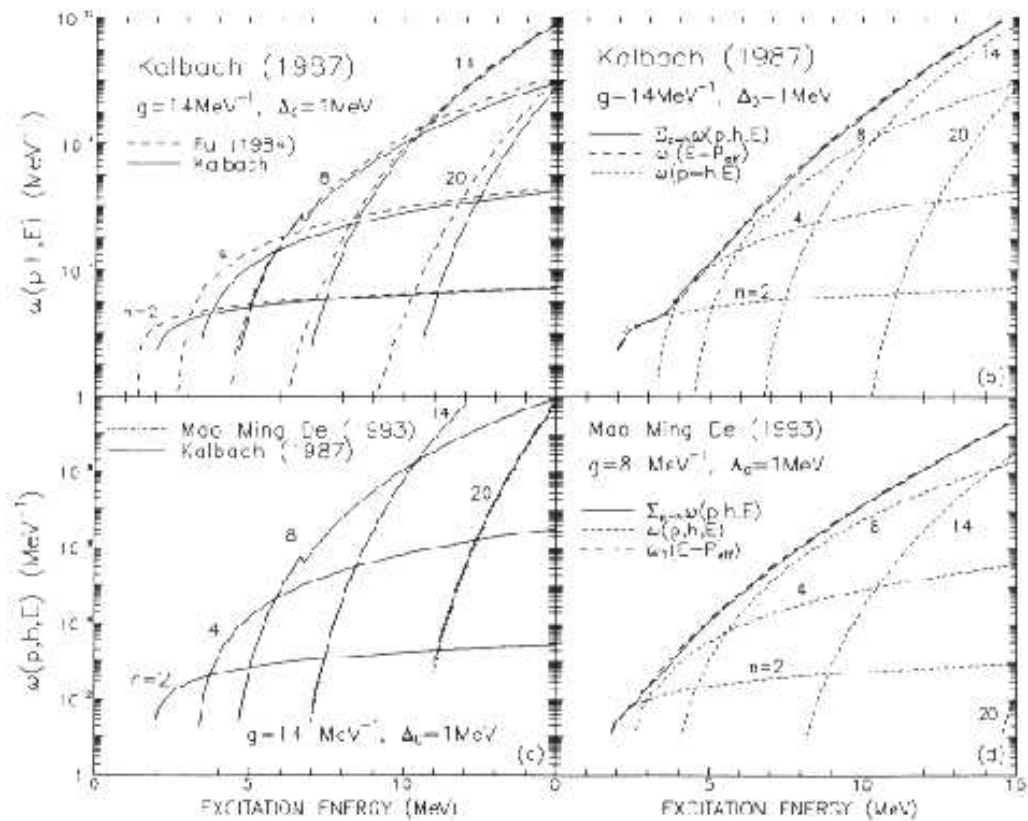
\item [FIG. 5.] Comparison of the PSDs obtained with the Williams
one-fermion formula, for the given exciton configurations, and the 
advanced pairing correction of (a) Fu \cite{fu84} (dashed curves) and 
Kalbach \cite{kalbach8789} above the threshold energy for each 
configuration (solid curves), as well as of (c) the latter and the 
PSDs with exact calculation of the Pauli-principle correction 
\cite{mao93} (dotted curves), (b) the sum of the Kalbach PSDs (shown
also below the threshold) for all allowed exciton numbers $p$=$h$ 
(solid curve) compared with the closed formula (dashed curve), for the 
values $g$=14 MeV$^{-1}$ and $\Delta_0$=1 MeV, and (d) the same but 
for the exact formula and the value $g$=8 MeV$^{-1}$ \cite{mao93}.

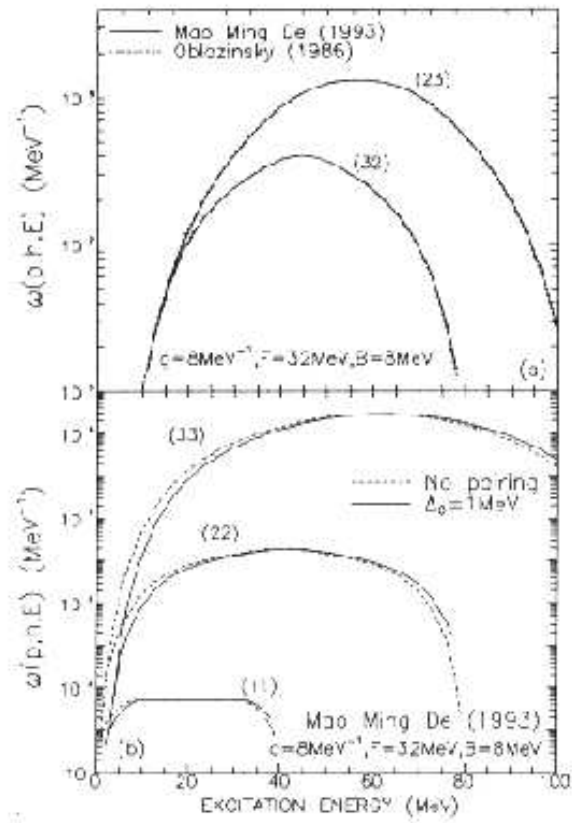
\item [FIG. 6.] Comparison of the particle-hole bound state 
densities for the given $ph$ configurations and the values $g$=8 
MeV$^{-1}$, $F$=32 MeV, and $B$=8 MeV, obtained with the Williams
one-fermion formula (a) including the nuclear-potential finite depth 
\cite{oblozinsky86} (dotted curves) and the exact calculation of the 
Pauli-principle correction \cite{mao93} (solid curves), as well as 
with (b) the latter form but without/with the advanced pairing 
correction and parameter value $\Delta_0$=1 MeV (dotted/solid curves).

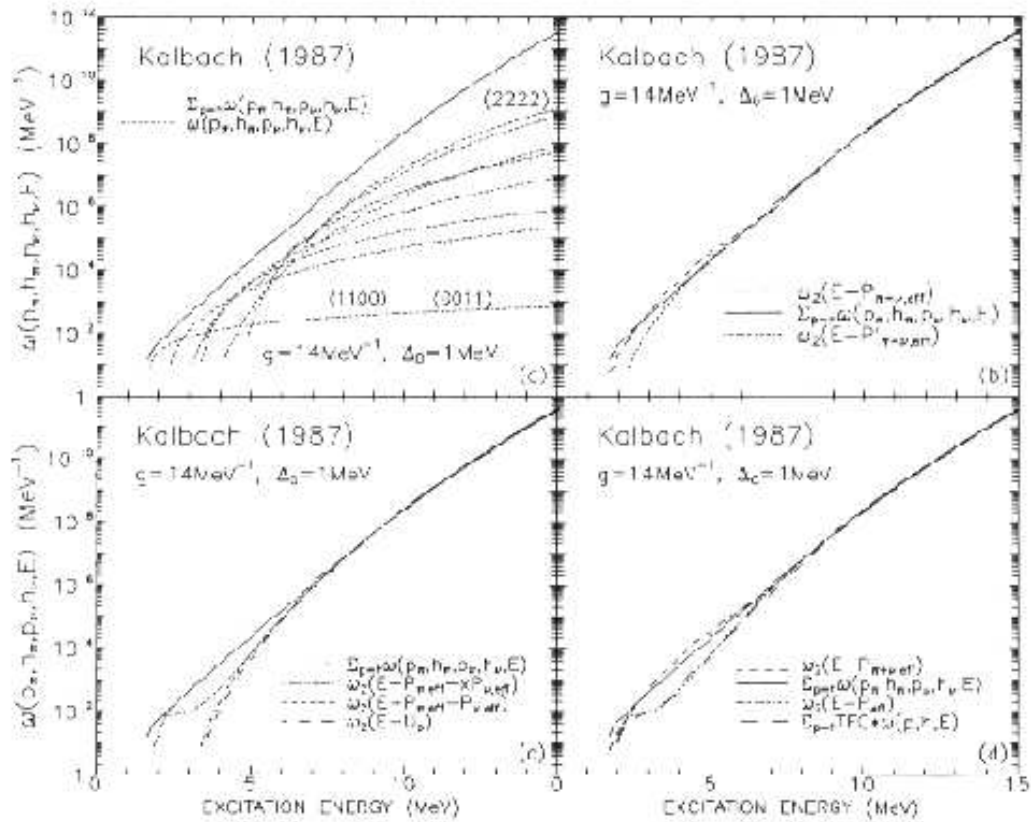
\item [FIG. 7.] The sum (solid curves) of (a) the two-fermion system 
PSDs \cite{kalbach8789} (dotted curves) for allowed exciton numbers 
$p_{\pi}$=$h_{\pi}$ and $p_{\nu}$=$h_{\nu}$, compared with (b) the 
closed formula of the two-component ESM total state density for the 
excitation energy decreased by the effective pairing correction given 
by the term $P'_{\pi+\nu,eff}$ (dotted curve), and (b),(d) the term 
$P_{\pi+\nu,eff}$ (dashed curve), (c) the constant $U_p$ (long-dashed 
curve), the sum $P_{eff}(E,C_{\pi})$+$xP_{eff}(E,C_{\nu})$ with $x$=1 
(dashed curve) as well as between 0 and 1 for $E/C$ varying from 1 to 
2 (dotted curve), and (d) the Kalbach term  $P_{eff}(E,C)$, also 
compared with the sum of the renormalized one-fermion PSDs 
\cite{akkermans85} (including the advanced pairing correction 
\cite{kalbach8789}) for all allowed exciton numbers $p$=$h$ 
(long-dashed curve), for the values $g$=14 MeV$^{-1}$ and 
$\Delta_0$=1 MeV.

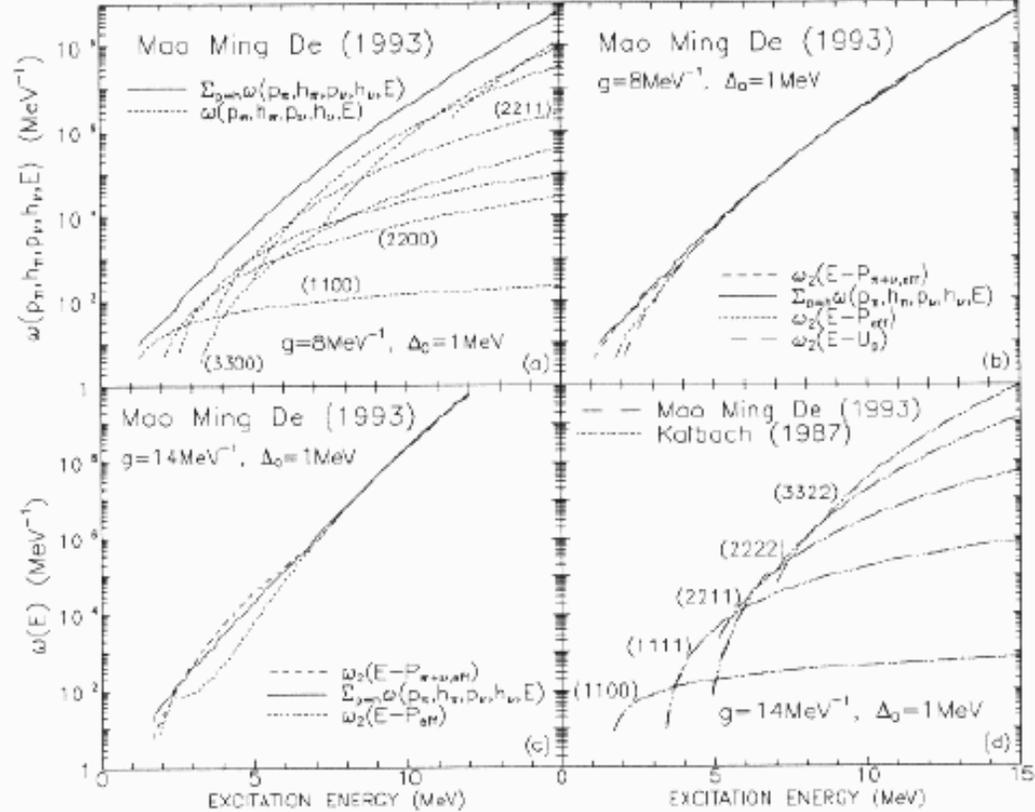
\item [FIG. 8.] The sum (solid curves) (a) of the two-fermion system 
PSDs \cite{mao93} (dotted curves) for allowed exciton numbers 
$p_{\pi}$=$h_{\pi}$ and $p_{\nu}$=$h_{\nu}$, compared with the closed 
formula of the two-component ESM total state density for the 
excitation energy decreased by the effective pairing correction given 
by (b),(c) the Kalbach effective-energy shift $P_{eff}(E,C)$ (dotted 
curve), the term $P_{\pi+\nu,eff}$ (dashed curve), and (c) the 
constant shift $U_p$. (d) The comparison of the PSD values of Refs. 
\cite{mao93} (dashed curves) and \cite{kalbach8789}, for the given 
exciton configurations. The global parameter values used are 
$\Delta_0$=1 MeV, and (a),(b) $g$=8 MeV$^{-1}$ 
\cite{oblozinsky86,mao93} and (c),(d) $g$=14 MeV$^{-1}$ 
\cite{fu84,kalbach8789}.

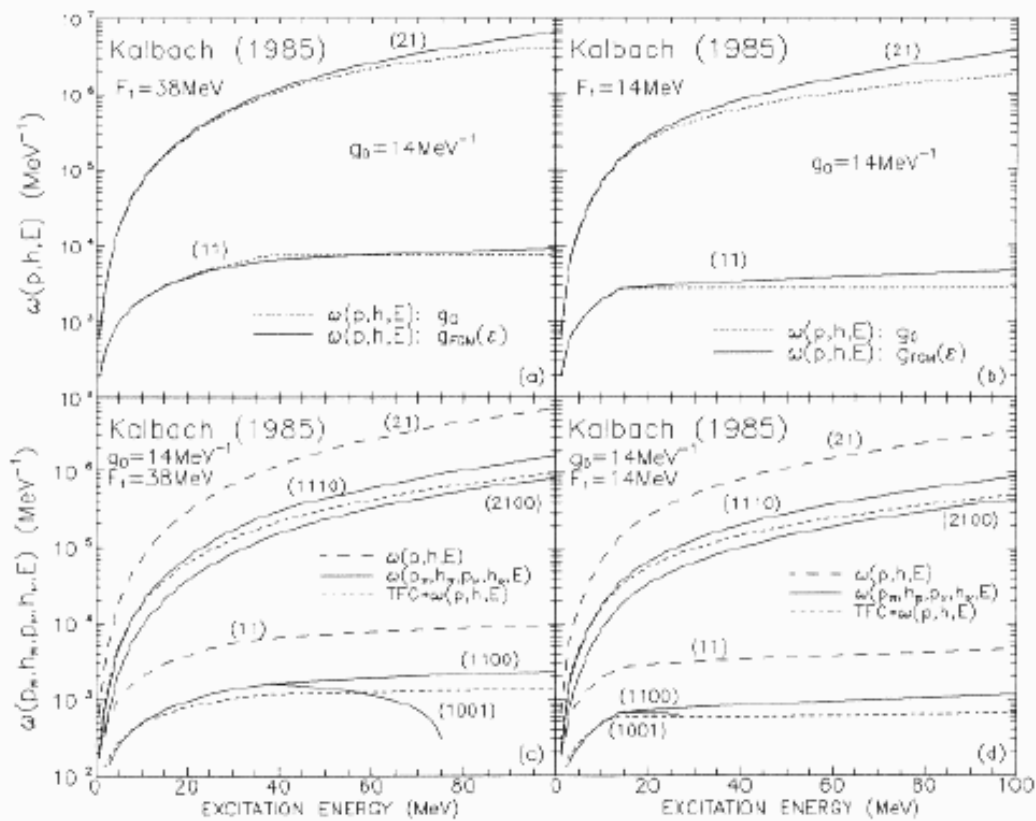
\item [FIG. 9.] PSDs for (a),(b) the given $ph$ configurations, 
obtained with the ESM one-fermion system formula with (solid dotted) 
and without (dotted curves) inclusion of the FGM energy dependence of 
the s.p.l. density, and (c),(d) the given 
$p_{\pi}h_{\pi}p_{\nu}h_{\nu}$ configurations, obtained with the 
two-fermion system formula including the FGM energy-dependent s.p.l.
density (solid curves) compared with the related results of the 
one-fermion system formula (dashed curves) and including the 
two-fermion system correction (dotted curves), as well as (a),(c) 
without and (b),(d) with the surface effect taken into account by 
using the average effective well depth $\overline{F}_1$=14  MeV 
\cite{kalbach85}, for the s.p.l. density value $g_0$=14 MeV$^{-1}$ 
and Fermi energy $F_0$=38 MeV.

\item [FIG. 10.] Comparison of the PSD values for the given $ph$
configurations and the sum providing the total state density, obtained 
by means of the given parameter values and (a),(b) the Williams 
(dotted curves) and the composite formula with no additional 
corrections (solid curves), for the one- and two-fermion systems 
respectively, and (c),(e) the Oblo\v{z}insk\'{y} one-fermion formula 
\cite{oblozinsky86} (dotted curves) and the composite formula 
including the finite-well depth and bound-state conditions (solid 
curves) as well as (d),(f) the average energy-dependent formalism. 

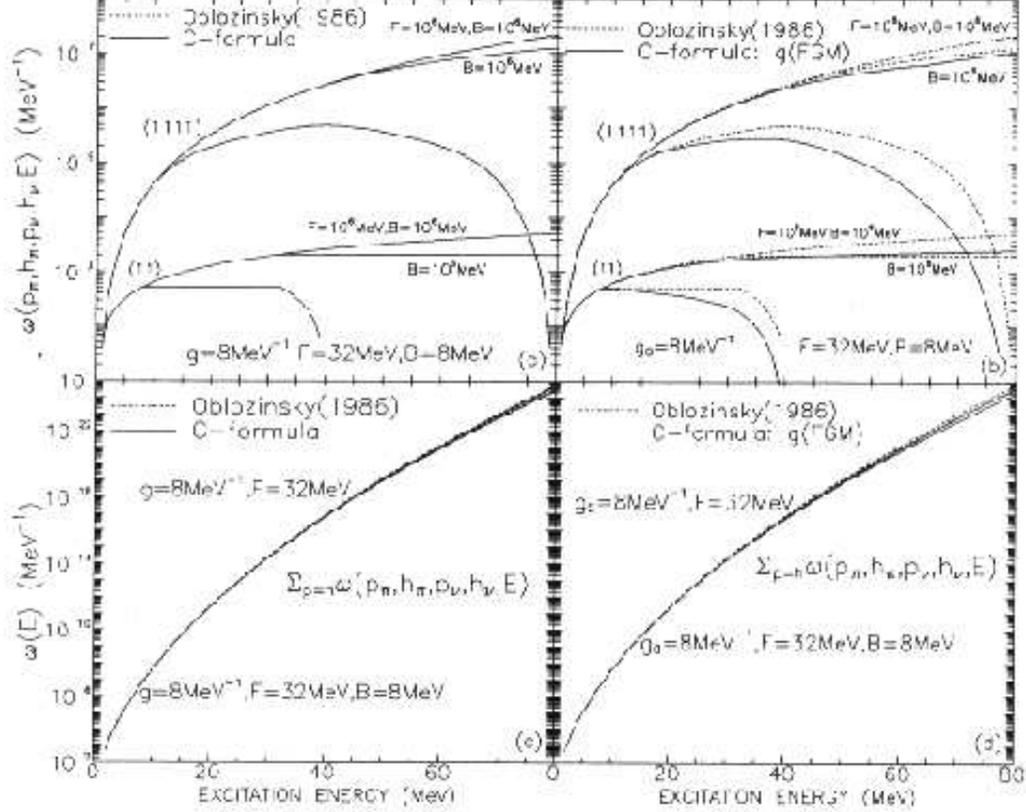
\item [FIG. 11.] Comparison of the PSD values for the given $ph$
configurations and the sum providing the total state density, obtained 
by means of the given parameter values and (a),(c) the 
Oblo\v{z}insk\'{y} one- and two-fermion system formulas 
\cite{oblozinsky86} (dotted curves) and the composite formula 
including the finite-well depth and bound-state conditions (solid 
curves) as well as additionally (b),(d) the energy-dependent s.p.l. 
densities within the composite formula. The PSDs for configurations 
including $1p1h$ which are obtained as noted in (a),(b) by releasing 
consecutively the finite-depth potential and bound-state conditions 
by means of large $F$- and $B$-values are also shown.

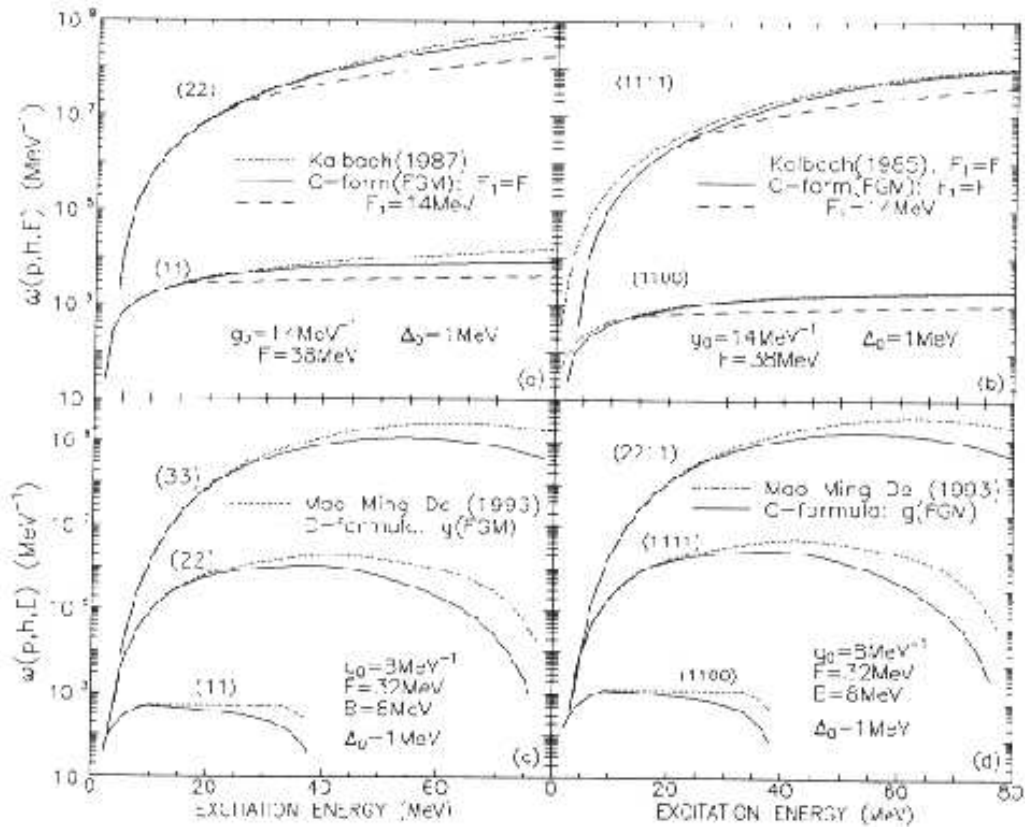
\item [FIG. 12.] Comparison of the PSD values for the given 
exciton configurations, obtained by means of the given parameter 
values and the composite formula including the finite-well depth, the 
energy-dependent s.p.l. densities (solid curves) and the surface 
effects taken into account by using the effective well depth 
$\overline{F}_1$=14 MeV (dashed curves), and (a) the Kalbach
one-fermion system formula \cite{kalbach8789} (dotted curves), (b)
the Kalbach two-fermion system formula \cite{kalbach85} (dotted 
curves), as well as (c),(d) for the bound states and the exact 
calculation of the Pauli-principle correction \cite{mao93} (dotted 
curves).

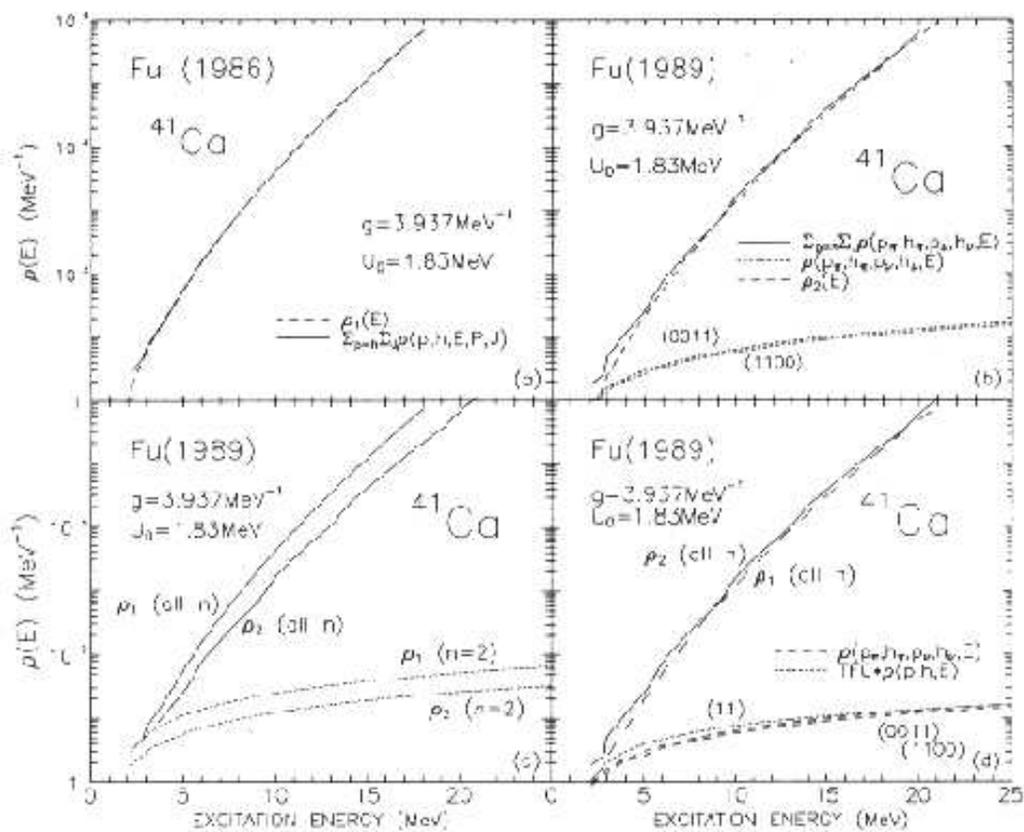
\item [FIG. 13.] Comparison of the total level density given by the 
sum of the ESM (a) one-fermion and (b) two-fermion system PLDs over 
both excited-particle number(s), equal to hole number(s), and spins 
(solid curves) and the related closed-formulas (dashed curves),
as well as of the closed-formula values obtained by using (c) the 
two-fermion system average spin-cutoff values $\sigma_2^2(n=2)$ 
\cite{fu91} and the correspondent $\sigma_1^2(n=2)$ for the 
one-fermion PLD (dotted curves) and the average values $\sigma_2^2$ 
and $\sigma_1^2$ for the related total level densities (solid curves 
noted "all n"), and (d) the two-fermion system formula (solid curve)
and the one-fermion formula with the two-fermion correction 
\cite{fu86,akkermans85} (long-dashed curve), for the excited nucleus 
$^{41}$Ca and the given parameter values. The corresponding PLD values 
are shown for configurations with two excitons (dotted curves).

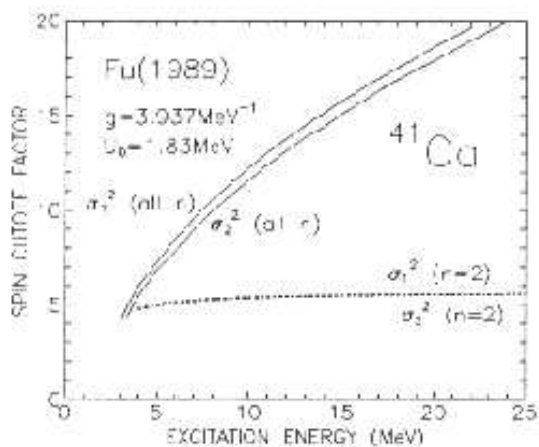
\item [FIG. 14.] Comparison of the two-fermion system average 
spin-cutoff values $\sigma_2^2(n=2)$ \cite{fu91} and the corresponding
$\sigma_1^2(n=2)$ for the one-fermion PLD (dotted curves), and the 
average values $\sigma_2^2$ and $\sigma_1^2$ (solid curves noted "all 
n"), for the excited nucleus $^{41}$Ca and the given parameter values.

\end{itemize}

\newpage
{\tighten
\begin{center}
  Table 1. Sample cases for partial state density (PSD) and partial 
level density (PLD) calculation up to excitation energy E$_{max}$.\\
\end{center}

\noindent
\bigskip
\begin{tabular}{lldd} \hline \hline
Case& \hspace*{1.in} TITLE (variable on first input record) &E$_{max}$&
								Execution\\
 No.&                      &(MeV)      &time$^a$ (s)\\ \hline  
1&Williams(1971)/Fig.2(one-fermion, asymptotical form in 2nd table!)&60&0.2\\
1A&Williams(1971) one-fermion formula, for distinct configurations&40&0.1\\
2&Williams(1971),as Fig.2 but two-fermion formula/asymptotical form&40&0.2\\
2A&Williams(1971) two-fermion formula, for distinct configurations&40&0.1\\
3&Williams(1971) one-fermion formula/asympt. form for Nb, g=A/13&35&0.1\\
3A&Williams(1971) two-fermion formula/asympt. form for Nb, g=A/13&35&0.3\\
3B&Williams(1971) one-fermion+TFC$^b$ formula/asympt.form for Nb, g=A/13&35&0.1\\
4&Betak+(1976)/Fig.1:one-fermion state density + finite depth corr.&400&0.3\\
5&Oblozinsky(1986)/Fig.1:one-fermion bound state dens. for configs.&80&0.1\\
5A&Oblozinsky(1986)/Fig.1: one-fermion state density + finite depth&80&0.1\\
5B&Oblozinsky(1986)/Fig.1: one-fermion state density formula&80&0.1\\
6&Oblozinsky(1986) one-fermion bound-state densities: g=8&80&0.4\\
6A&Oblozinsky(1986) one-fermion state density + finite depth corr.&80&0.4\\
6B&Oblozinsky(1986) one-fermion formula/asymptotical form: g=8&80&0.3\\
7&Oblozinsky(1986) two-fermion bound-state densities: g=8&80&9.2\\
7A&Oblozinsky(1986) two-fermion state density +finite depth corr., g=8&80&7.0\\
7B&Oblozinsky(1986) two-fermion formula/asymptotical form: g=8&80&6.8\\
8&Fu(1984)/Fig.3: one-fermion state density with/without pairing&15&0.3\\
9&Fu(1984)/Fig.6: one-f. closed-formula values for various pairings&20&0.3\\
10&Fu(1984)/Fig.7: one/two-f. closed-formula with various BSFG data&40&0.2\\
10A&Fu(1984)/Fig.7: two-f. closed-formula with two-f. correction&40&0.1\\
11&Fu(1986)/Fig.6:one-fermion partial level dens. sum/closed-formula&11&0.1\\
11A&Fu(1986) one-fermion partial level density, total PLD for config.&11&0.1\\
11B& Fu(1986)/Fig.1: two-fermion PLD for 41Ca: sum/closed-formula&25&0.2\\
11C&Fu(1989)/Fig.1: two-fermion PLD for 41Ca configuration &25&0.1\\
11D&Fu(1989)/Fig.1: two-fermion PLD for 41Ca /sum/closed-formula &25&0.1\\
11E&Fu(1989)/Fig.1: one-fermion PLD + TFC: sum/closed-formula &25&0.1\\
12&Kalbach(1987)/Fig.3: one-fermion partial state dens. for configs. &15&0.1\\
13&Kalbach(1987) one-fermion partial state density/sum/closed form&15&0.1\\
13A&Kalbach(1987) two-fermion PSD sum/closed formula,for g=8/Up=3.5 &15&0.3\\
13B&Kalbach(1987) one-fermion PSD + TFC /sum/closed form: g=8/Up=3.5 &15&0.1\\
14&Kalbach(1989)/Tab.1: one-fermion partial state dens. for configs.&32&0.1\\
14A&Kalbach(1989)/Tab.1b:one-fermion partial state dens. for configs.&128&0.1\\
14B&Kalbach(1989)/Tab.1c:one-fermion partial state dens. for configs.&256&0.1\\
15&Mao Ming De(1993)/Fig.1:one-fermion bound PSD + exact Pauli corr. &100&106.7\\
15A&Mao Ming De(1993)/Fig.2/Oblozinsky(86):bound PSD+exact Pauli corr.&100&42.9\\ 
16&Mao Ming De(1993)/Fig.3:bound PSD+exact Pauli corr.+pairing corr. &100&64.1\\
17&Mao Ming De(1993):one-fermion PSDs+pairing corr./sum/closed form&15&123.4\\
17A&Mao Ming De(1993)/Kalbach(87):one-f. PSDs: exact Pauli corr. effect&15&134.6\\
17B&Mao Ming De(1993):two-fermion PSDs+pairing corr./sum/closed form &15&351.5\\
17C&Mao Ming De(1993):1-f.PSDs+pair.corr.(2-3.5)+TFC/sum/closed form &15&257.4\\
17D&Mao Ming De(1993)/Kalbach(87):two-f. PSDs: exact Pauli corr. effect &15&427.1\\
\hline
\end{tabular}
\newpage
\hspace*{1.7in}Table 1. -- continued\\

\noindent
\begin{tabular}{lldd} \hline 
Case& \hspace*{1.in} TITLE (variable on first input record) &E$_{max}$&
								Execution\\
 No.&                      &(MeV)      &time$^a$ (s)\\ \hline  
18&Kalbach(1985)/(Fig.2):one-fermion PSD with Finite Depth Cor.:F1=38&100&0.1\\
18A&Kalbach(1985)/(Fig.2):one-f. Energy-dep. s.p.s. density+FDC$^c$:F=38&100&0.1\\
18B&Kalbach(1985)/(Fig.2):one-f. PSD with FDC+surface effects: F1=14&100&0.1\\
18C&Kalbach(1985)/(Fig.2):E-dep. s.p.s.density+FDC:F=38+surf.ef.:F1=14&100&0.1\\
19 &Kalbach(1985)/(Fig.2): two-fermion PSD + FDC (F=38), g-FGM for &100&0.2\\
19A&Kalbach(1985): one-fermion PSD + FDC (F=38) + TFC, for configs.&100&0.1\\
20 &Kalbach(1985)/(Fig.2): two-fermion PLD + FDC (F1=14), g-FGM, for &100&0.2\\
20A&Kalbach(1985): one-fermion PLD + FDC (F=14) + TFC &100&0.1\\
21 &C-formula/Williams(1971)/Fig.2: one-fermion PSD /sum/closed form &80&0.3\\
21A&C-formula/Williams(1971)/Fig.2: one-fermion PSD for configs.&60&0.1\\
22 &C-formula/Williams(1971): two-fermion PSD /sum/closed formula &40&0.5\\
22A&C-formula/Williams(1971): two-fermion PSD for distinct configs.&40&0.1\\
23 &C-formula/Oblozinsky(1986)/Fig.1: one-f. bound PSD for configs.&80&0.1\\
23A&C-formula/Oblozinsky(1986): one-fermion bound PSD/sum/closed form &80&0.11\\
23B&C-formula/Oblozinsky(1986): two-fermion bound PSD/sum/closed form &80&24.8\\
23C&C-formula(g-FGM)/Oblozinsky(1986): one-f. bound PSD for configs.& 80&3.5\\
23D&C-formula(g-FGM)/Oblozinsky(1986): 2-f. bound PSD/sum/closed form &80&28.1\\
24 &C-formula/Oblozinsky(1986): one-fermion PSD + FDC for configs.&80&0.1\\
24A&C-formula/Oblozinsky(1986): one-f. PSD + FDC /sum/closed form &80&1.0\\
24B&C-formula/Oblozinsky(1986): two-f. PSD + FDC /sum/closed form &80&17.3\\
24C&C-formula(g-FGM)/Oblozinsky(1986): one-f. PSD+FDC/sum/closed form&80&1.5\\
24D&C-formula(g-FGM)/Oblozinsky(1986): two-f. PSD+FDC/sum/closed form&80&19.7\\
25 &C-formula/Oblozinsky(1986): one-fermion ESM PSD for configuration &80&0.1\\
25A&C-formula/Oblozinsky(1986): one-fermion ESM PSD/sum/closed form &80&1.0\\
25B&C-formula/Oblozinsky(1986): two-fermion ESM PSD /sum/closed form &80&16.7\\
26 &C-form(g-FGM)/Kalbach(1987):one-f.PSD+FDC(F=38MeV)/sum/closed form&80&2.3\\
26A&C-form(g-FGM)/Kalbach(1987):1-f.PSD+FDC(F1=14MeV)/sum/closed form&80&2.5\\
27 &C-form(g-FGM)/Kalbach(1985):two-f.PSD+FDC(F=38MeV)/sum/closed form&80&46.7\\
27A&C-form(g-FGM)/Kalbach(1985):2-f.PSD+FDC(F1=14MeV)/sum/closed form&80&46.9\\
28 &C-form(g-FGM)/Mao Ming(1993): 1-f.PSD+FDC(F=32MeV)/sum/closed form&80&276.6\\
28A&C-form(g-FGM)/Mao Ming(1993):2-f.PSD+FDC(F=32MeV)/sum/closed form&80&1679.3\\
\hline \hline
\end{tabular}
\\
$^a$ For calculation carried out on PC {\it Pentium/166MHZ Intel}.\\
$^b$ Two-fermion system correction (TFC).\\
$^c$ Nuclear potential finite-depth correction (FDC).

\newpage
\begin{center}
  Table 2. Input data in reading order.\\
\end{center}

\small
\noindent
\begin{tabular}{clll} \hline \hline
Record&Variable&FORMAT& Meaning \\ \hline  
1 &  NE &2I3,74A1& if NE$<$0, $\vert$NE$\vert$ is the number of 
			excitation energies to be read ($\le$200);\\ 
  &     &        & if NE$>$0, the excitation energies are 
					(FLOAT(I),I=1,NE) (MeV)\\
  &IOPTJ&        & spin distribution (PLD calculation) option:\\
  &     &        & = 0, $\omega(p,h,E)$ or 
      		    $\omega(p_{\pi},h_{\pi},p_{\nu},h_{\nu},E)$ are calculated\\
  &     &        & = 1, $\rho(p,h,E,J)$ or 
		    $\rho(p_{\pi},h_{\pi},p_{\nu},h_{\nu},E,J)$ are calculated\\
  &TITLE&        & title of the problem\\
2 &(E(I),I=1,-NE) & 8F10.5 & [record to be read only if NE$\le$0]\\
  &               &        & excitation energies (MeV) for PSD/PLDs
					 calculation\\
3 &IMOD &2I3,7F10.5& option for PSD formula (odd/even: one/two-fermion system
			 formulas):\\
  &     &          & =-1, 0: composite (recommended) formula\\
  &     &          & = 1, 2: F.C. Williams, Nucl. Phys. A166, 231 (1971)\\
  &     &          & = 3, 4:  P.Oblozinsky, Nucl. Phys. A453, 127 (1986), 
				Eqs.(7,9)\\
  &     &          & = 5, 6:  C.Y. Fu, Nucl. Sci. Eng. 86, 344 (1984)\\
  &     &          & = 7, 8:  C. Kalbach, Nucl.Sci.Eng.95,70(1987), 
				Z.Phys.A 332,157(1989)\\
  &     &          & = 9,10:  C. Kalbach, Phys. Rev. C 32, 1157 (1985)\\
  &     &          & =11,12:  Mao Ming De Gua Hua, J. Phys. G 19, 421 (1993)\\
  &ITFC &          & option for the two-fermion system correction:\\
  &     &          & = 0: no two-fermion system correction\\
  &     &          & = 1: J.M.Akkermans, H.Gruppelaar, Z. Phys. A 321,605(1985), 
			Eq. (9)\\  
  &  A  &          & excited-nucleus mass number (it may be omitted if 
			GIN$\ge$0.)\\
  &  Z  &          & excited-nucleus atomic number\\
  & UP  &          & pairing correction based on the odd-even mass differences\\
  &     &          & (if UP=-1. and A$\ge$0., Eq.(9) of \cite{dilg73} is 
			adopted)\\
4 & NP0 &2I3,7F10.5& [record to be read only for odd IMOD]
				number $p$ of excited particles\\
  & NH0 &          & number $h$ of holes\\
  &     &          & (if NP0=NH0=0, calculation is done for all pairs 
			$p$=$h$=1, 2,... \\
  &     &          &    for which PSD/PLD-value is $\ge$0.1 MeV$^{-1}$)\\
  & GIN &          & single-particle state density G (MeV$^{-1}$):\\
  &     &          & if GIN$\le$0., then G=(6/3.14$^2$2)*DR(1) is adopted;\\
  &     &          & if GIN=0 and A=0, then G=1.0 is adopted;\\
  &     &          & if GIN=0 and A$\ge$0, then G=A/13.0 is adopted\\
  & FIN &          & Fermi energy F (MeV); if FIN=0, then F=10$^6$ is adopted\\
  & BIN &          & nucleon binding energy B (MeV); if BIN=0, then B=10$^6$ 
			is adopted\\
  &F1IN &          & average effective Fermi energy F1 (MeV); if FIN=0, then
			F1=10$^6$;\\
  &     &          & if FIN$\le$0, then constant G is used within the WR1 and WR2 
			functions\\
\hline \hline
\end{tabular}
\newpage
\hspace*{1.7in}Table 2. -- continued\\

\noindent
\begin{tabular}{clll} \hline
Record&Variable&FORMAT& Meaning \\ \hline  
5 & NP0 &4I3,6F10.5& [record to be read only for even IMOD]
			number $p_{\pi}$ of proton excited particles\\
  & NH0 &          & number $h_{\pi}$ of proton holes\\
  & NPN0&          & number $p_{\nu}$ of neutron excited particles\\
  & NHN0&          & number $h_{\nu}$ of neutron holes\\
  &     &          & (if NP0=NH0=NPN0=NHN0=0, all configurations
		      ($p_{\pi}$=$h_{\pi}$,$p_{\nu}$=$h_{\nu}$)\\ 
  &     &          & with N=2, 4, .. and PSD/PLD sum $>$0.1 MeV$^{-1}$ are 
			considered)\\
  & GIN &          & single-proton state density G (MeV$^{-1}$):\\
  & GN  &          & single-neutron state density GN (MeV$^{-1}$):\\
  &     &          & if GIN$\le$0., then G=Z/A*(6/3.14$^2$2)*DR(1) (see record 6)\\
  &     &          & \hspace*{0.4in} and GN=(A-Z)/A*(6/3.14$^2$2)*DR(1) are adopted;\\
  &     &          & if GIN=0 and A=0, then G=GN=1.0 are adopted;\\
  &     &          & if GIN=0 and A$\ge$0, then G=Z/13.0 and GN=(A-Z)/13.0
			are adopted\\
  & FIN &          & proton Fermi energy F (MeV); \\
  & FN  &          & neutron Fermi energy FN (MeV); \\
  &     &          & if FIN=0, then F=FN=10$^6$ are adopted\\
  & BIN &          & proton binding energy B (MeV); if BIN=0, then B=10$^6$
			is adopted\\
  & BN  &          & neutron binding energy B (MeV);\\
  &     &          & if BIN=0, then B=BN=10$^6$ are adopted\\
  &F1IN &          & average effective proton Fermi energy F1 (MeV); \\
  &     &          & if F1IN=0, it is adopted the value F1=10$^6$;\\
  &     &          & if F1IN$\le$0, constant G is used within WR1 and WR2 
			functions\\
  & F1N &          & average effective neutron Fermi energy F1N (MeV); \\
  &     &          & if F1N=0, it is adopted the value F1N=10$^6$;\\
  &     &          & if F1N$<$0, constant GN is used within WR1 and WR2 
			functions\\
6 &DR(I)& 8F10.5& [record to be read only if GIN$\le$0] 
			BSFG level density parameters \cite{dilg73}:\\
  &             &      & DR(1) - nuclear level density parameter $a$;\\
  &             &      & DR(2) - ratio of effective nuclear moment of inertia 
				to rigid-body value;\\
  &             &      & DR(3) - shift of the fictive nuclear ground state\\
7 &ICONT&2I3& output and recycle option:\\
  &     &   & =-1, print first 2 tables of calculated PSD/PLDs (see PRINTWN 
			description);\\
  &     &   & \hspace*{0.3in} resumption according to IEND;\\
  &     &   & = 0, print calculated PSD/PLDs; resumption according to IEND;\\
  &     &   & = 1, new case by input-data record 1; results printed 
			at once with previous case;\\
  &     &   & = 2, new case by input-data record 3 (same energy grid);\\
  &     &   & = 3, calculation for other exciton configuration by input-data 
			record 4 or 5;\\
  &     &   & = 4, calculation for other BSFG parameter set by input-data 
			record 6\\
  &IEND &   & recycle option:\\
  &     &   & = 0, end;\\
  &     &   & = 1, new complete case by input-data record 1;\\
  &     &   & = 2, new case by input-data record 3;\\
  &     &   & = 3, new case for exciton configuration by record 4 or 5;\\
  &     &   & = 4, new case for BSFG parameters by record 6\\
\hline \hline
\end{tabular}
\\
\normalsize}
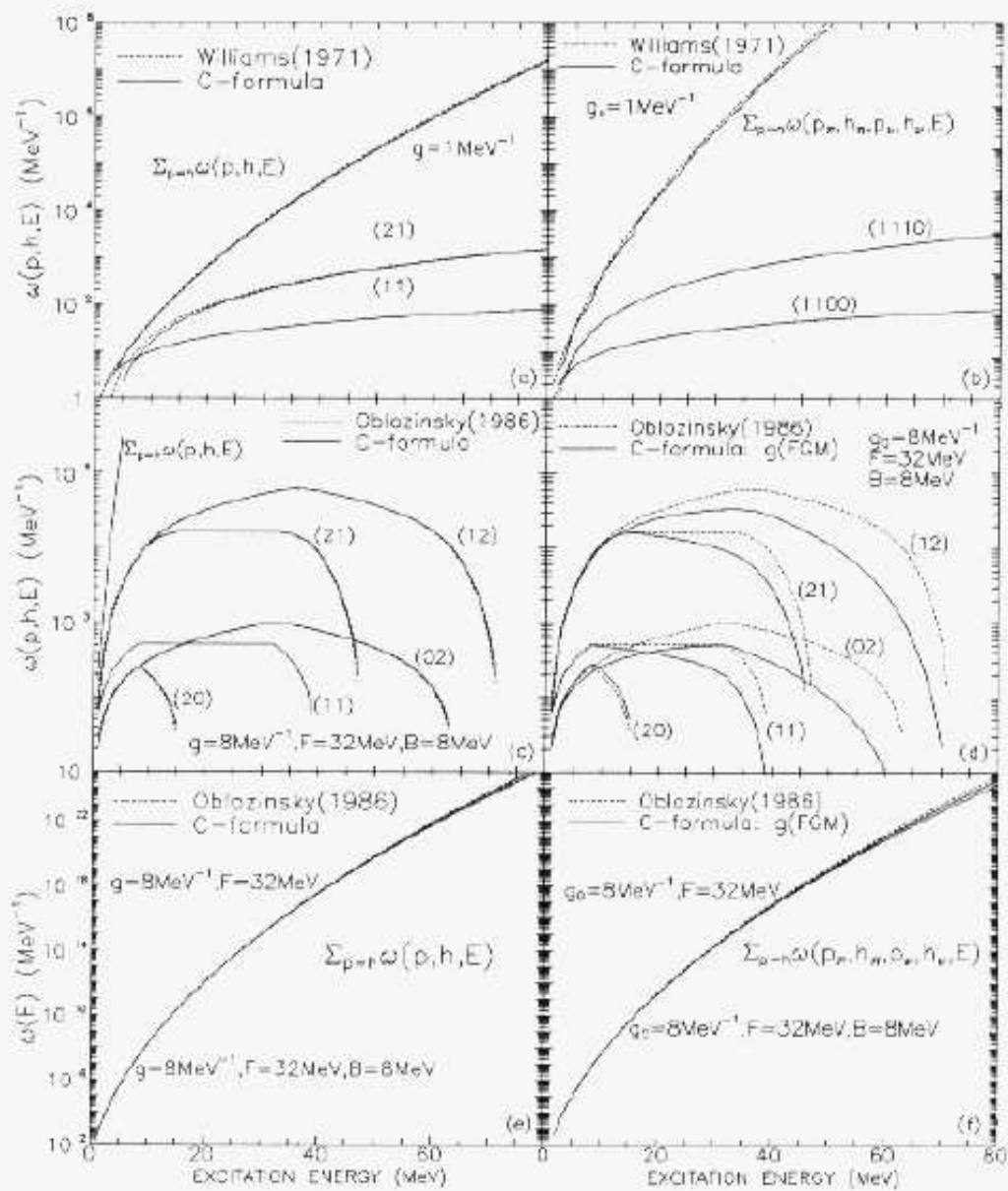
\end{document}